\documentclass[preprint,aps,amsmath,amssymb]{revtex4-1}
\usepackage[utf8]{inputenc}
\pdfoutput=2

\usepackage{graphicx}
\usepackage{dcolumn}
\usepackage{bm}
\usepackage{pdfsync}
\usepackage{xfrac}
\usepackage{amsmath}
\usepackage[per-mode=symbol-or-fraction,separate-uncertainty=true]{siunitx}
\DeclareSIUnit{\kgpcm}{\kilogram\per\cubic\metre}
\DeclareSIUnit{\gpcc}{\gram\per\cubic\centi\metre}
\newcommand{\atompair}[2]{#1-#2}
\newcommand{\lj}[0]{\mbox{L-J}}
\newcommand{\ljbaseline}{\lj\ baseline}

\newcommand{\abinit}[0]{\textit{ab initio}}
\newcommand{\etal}[0]{\textit{et al.}}
\newcommand{\bvec}[1]{\ensuremath{\mathbf{#1}}}

\usepackage[disable]{todonotes}

\usepackage[
]{geometry}

\begin{document}

\title{Equation of state of fluid methane from first principles\\
with machine learning potentials}

\author{Max Veit}
\email{max.veit@epfl.ch}
\altaffiliation{
Current address:
Laboratory of Computational Science and Modeling,
Ecole Polytechnique Fédérale de Lausanne,
1015 Lausanne, Switzerland
}
\affiliation{
Engineering Laboratory\\
University of Cambridge\\
Trumpington Street\\
Cambridge, CB2 1PZ\\
United Kingdom
}

\author{Sandeep Kumar Jain}
\author{Satyanarayana Bonakala}
\author{Indranil Rudra}
\affiliation{
Shell India Markets Pvt. Ltd.\\
Bengaluru 562149\\
Karnataka, India
}

\author{Detlef Hohl}
\affiliation{
Shell Global Solutions International BV\\
Grasweg 31\\
1031 HW Amsterdam\\
The Netherlands
}

\author{Gábor Csányi}%
\affiliation{
Engineering Laboratory\\
University of Cambridge\\
Trumpington Street\\
Cambridge, CB2 1PZ\\
United Kingdom
}

\date{\today}




\begin{abstract}
    The predictive simulation of molecular liquids requires models that are not
    only accurate, but computationally efficient enough to handle the large
    systems and long time scales required for reliable prediction of
    macroscopic properties.  We present a new approach to the systematic
    approximation of the first-principles potential energy surface (PES) of
    molecular liquids using the GAP (Gaussian Approximation Potential)
    framework.  The approach allows us to create potentials at several
    different levels of accuracy in reproducing the true PES, which
    allows us to test the level of quantum chemistry that is necessary to accurately
    predict its macroscopic properties.  We test the approach by building
    potentials for liquid methane (CH\textsubscript{4}), which is difficult to
    model from first principles because its behavior is dominated by
    weak dispersion interactions with a significant many-body component.  We
    find that an accurate, consistent prediction of its bulk density across a
    wide range of temperature and pressure requires not only many-body
    dispersion, but also quantum nuclear effects to be modeled accurately.
\end{abstract}

\maketitle

\section{Introduction}
The accurate simulation of molecular liquids is a problem of great scientific
and industrial importance. We not only would like to be able to test the predictions of our models
against experimental benchmarks to see where they need to be refined, but
we also need to make predictions for new compounds or mixtures in order to
identify the most promising candidates for future applications. When modeling molecular liquids one is typically obliged
to trade off accuracy in the description of the potential energy surface and errors due
to insufficient sampling. Here we show how one can use function fitting techniques analogous 
to those in machine learning to create models that reach quantum chemistry accuracy at a much
reduced cost. We also break down the total interaction potential into different components
that allow us to show explicitly that all components have been modeled sufficiently accurately,
and thus we obtain the right answers {\em for the right reasons} rather than due to uncontrolled cancellation
of errors.

In this work
we aim to perform simulations of \textit{ab
initio} quality but with the orders of magnitude boost in computational
efficiency afforded by high dimensional regression using machine learning.  We create
Gaussian approximation potentials (GAPs)~\cite{gap-prl,Bartok2015a,Gillan2013}
for liquid methane, the simplest alkane, which is inherently difficult to model
because its behavior is dominated by weak dispersion interactions.  It is also
useful as a stepping stone towards potentials that can model larger hydrocarbons
under more extreme conditions~\cite{Kaminski1994,Hayes2004}; such a potential
would enable new research in numerous scientific and engineering
applications~\cite{Spanu2011,Payal2012,Hansen1999}.

There is a long history of modeling liquids at the atomistic scale with Monte
Carlo (MC) or molecular dynamics (MD) methods.  The interactions between
constituent particles are often modeled using analytical potentials, which are a
combination of a few simple, physically motivated expressions, such as the
venerable Lennard-Jones potential~\cite{Lennard-Jones1931} (the 12-6 form,
hereafter referred to as \lj) and the many subsequent variations or extensions
of this basic
form~\cite{Buckingham1947,Weiner1986,Jorgensen1984,Sun1998,Stuart2000,Allen+1989}.
These potentials contain empirical parameters which are usually optimized until the
simulations reproduce specific sections of the experimental equation of state.

\begin{figure}[p]
    \centering
\includegraphics[scale=0.75]{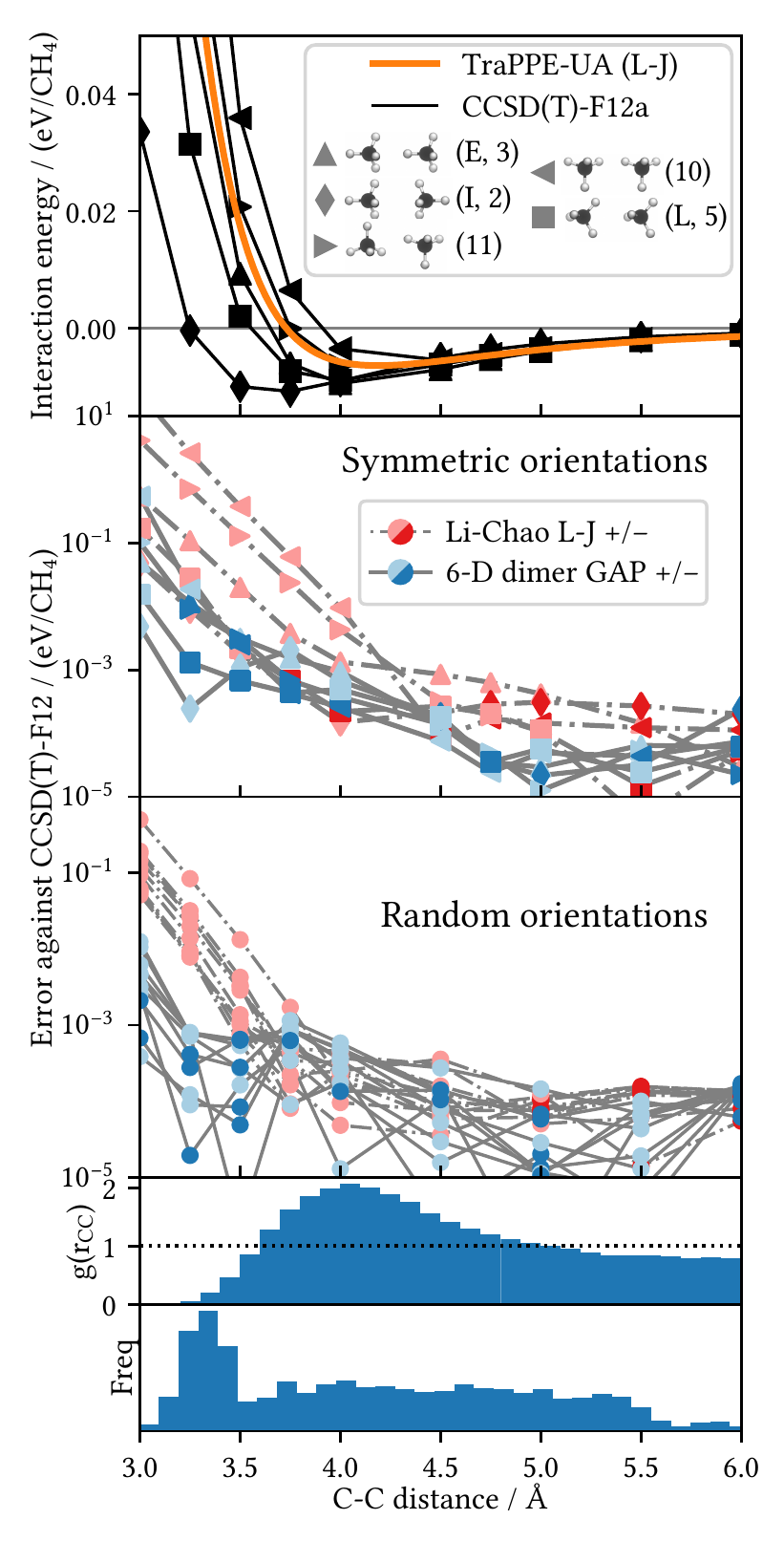}
\caption{Top: Interaction
    energies of the rigid methane dimer in a selection of orientations.  The
    TraPPE united-atom (and therefore isotropic) model~\cite{Martin1998} is given by the smooth line.
    Configurations are labeled as in Chao \etal~\cite{Chao2009} (letters) and
    Hellmann \etal~\cite{Hellmann2008a} (numbers).  Middle: Errors of two models
    of the methane dimer energy, the pairwise \lj\ fit of Li and
    Chao~\cite{Li2016} and a full-dimensional GAP fit, against CCSD(T)-F12 on
    the same orientations.
    Bottom: Errors with ten randomly chosen orientations.  A pair correlation
    function at \SI{188}{\kelvin} and \SI{278}{\bar} and a histogram of the fitting
    database are given below for reference.  Dimer pictures made with VMD~\cite{HUMP96}.
}
\label{fig:dimer-anisotropy}
\end{figure}

Recent potentials show a trend of more closely representing the underlying
quantum mechanical potential energy surface, for example by adding
anharmonic and cross terms to the covalent forces to arrive at a more faithful
representation~\cite{Maple1994,Allinger1996,Sun1998}  or even directly fitting
the intramolecular~\cite{Allen2018} or
intermolecular~\cite{Gay1991,Jalkanen2003,Hellmann2008a,Hayes2004,Li2016} terms
to \abinit\ calculations.
However, the analytical parameterized functional forms employed in such
potentials remain too simple to represent the underlying potential energy
surface faithfully; Figure~\ref{fig:dimer-anisotropy} shows that a pair
potential~\cite{Li2016} fit directly to coupled-cluster data cannot represent
the complex, anisotropic potential energy surface of the methane dimer.  It
additionally shows a potential that we fit to CCSD(T) using the Gaussian Approximation
Potentials (GAP) method~\cite{gap-prl,Bartok2015a} (more details in the
supporting information) in the full six-dimensional space of mutual dimer
orientations (monomers kept rigid); this potential, which we will call the `6-D
dimer GAP', is indeed capable of representing the complex anisotropy of the
dimer's potential energy surface.  Hence, potentials fit with fixed functional
forms must represent thermal averages of this surface that are useful for making
predictions within a certain range of temperature and pressure; these
predictions typically break down once the simulations are taken outside of this
range (see Figure~\ref{fig:isotherms}).  It is indicative and somewhat sobering
that the most accurate prediction of the density of liquid methane is achieved
by the {\em simplest} potentials, TraPPE (the coarse united atom version
TraPPE-UA~\cite{Martin1998} and the reduced dimensional version
TraPPE-EH~\cite{Chen1999}). Neither attempts to
reproduce the actual Born-Oppenheimer potential energy surface; in fact, every
effort up to now to capture the potential energy surface by an analytical
potential has lead to worse predictions of the liquid density. 

Several methods are available that approximate the true quantum potential energy
surface.  Perhaps the best known of these is density functional theory
(DFT)~\cite{Martin+2008}, which is generally good at predicting covalent bond
energies and intermolecular repulsive interactions.
Standard DFT lacks dispersion interactions, however, so these must be added
separately~\cite{Grimme2016}.  Dispersion correction schemes for DFT are
generally inverse-power terms added on to the total DFT energy.  They range from
terms with fixed semiempirical coefficients~\cite{Grimme2006} to explicitly
geometry-dependent terms~\cite{Grimme2010}, to terms with coefficients that use
information from an existing DFT
calculation~\cite{Becke2006,Tkatchenko2009,Tkatchenko2012}.  Many of these
schemes, such as DFT-D3~\cite{Grimme2010} and MBD~\cite{Tkatchenko2012}, account
for many-body (i.e. beyond pairwise additive) dispersion interactions.  This
many-body effect has been shown to be crucial for an accurate description of
many dispersion-bound systems such as supramolecular
complexes~\cite{Blood-Forsythe2016} and organic crystals~\cite{Shtukenberg2017},
though the effects on molecular liquids have not yet been extensively studied --
a many-body vdW model (D3~\cite{Grimme2010}) \emph{was} included in the water
potential of Morawietz \etal~\cite{Morawietz2016}, but it was not mentioned
whether a simple pairwise model would have given different results.

The main drawback of quantum methods that treat electrons explicitly, such as DFT
or quantum chemistry, is their computational cost: MD simulations to predict
liquid properties routinely require millions of force evaluations on thousands
of atoms~\cite{Payal2012}, which would be prohibitive even for today's fastest
computers using the most efficient implementations of DFT.  Furthermore, MD
simulations require force evaluations on many highly correlated configurations.
But the Born-Oppenheimer potential energy surface is usually assumed to be
smooth and regular, at least in the ordinary realm of closed-shell molecules far
away from level crossings and other exotic PES irregularities.  Thanks to this
regularity, highly correlated (similar) configurations will also have highly
correlated energies and forces.  This correlation can be exploited to greatly
reduce the number of force evaluations required for a molecular simulation.

\subsection{Machine learning potentials}
A new generation of potentials aims to
exploit this correlation by using machine learning techniques to directly fit
the Born-Oppenheimer potential energy surface~\cite{gap-prl,Behler2007}.  These
fits do not constrain the potential's functional form, relying instead on a
sufficient sample of existing calculations to be able to regress (fit) these
data points in the high-dimensional space of nuclear positions.  Such potentials
are designed to capture much of the accuracy and flexibility offered by full
quantum methods but with a computational efficiency that is many orders of
magnitude higher, enabling MD simulations for system sizes and timescales
previously only accessible to empirical, analytical potentials.

Machine learning potentials have been applied to a wide variety of
systems~\cite{jcp-editorial-datachem}.  In our group, they have been applied to
systems ranging from solids such as silicon~\cite{gap-prl},
tungsten~\cite{Szlachta2014}, iron~\cite{Dragoni2018}, and
boron~\cite{Deringer2018}, molecular clusters~\cite{Gillan2013} and
liquids~\cite{Bartok2013,Morawietz2016}, and amorphous
materials~\cite{Deringer2017}.  There is also considerable interest in general,
transferable molecular potentials~\cite{Bereau2015} and accurate modeling of
liquid water~\cite{Morawietz2016}.  Recent progress has also been made in
modeling across different chemical compounds~\cite{De2016,Montavon2013} and even
across different classes of materials~\cite{Bartok2017}, thus approaching the
level of flexibility currently offered by full quantum methods.

\subsection{Quantum nuclear effects}
Empirical potentials have been fit to reproduce experimental equations of state,
so they include quantum nuclear effects implicitly.  In contrast, when simulations
are done with a systematic approximation of the Born-Oppenheimer potential
energy surface, it becomes necessary to account for quantum nuclear effects in
an equally systematic manner~\cite{Markland2018,Kapil2016}.  These effects are
especially important at low temperatures and with light nuclei; their
importance in liquid alkanes in particular has long been
established~\cite{Balog2000} and was recently
highlighted~\cite{Pereyaslavets2018} using quantum mechanically fitted
forcefields.  In empirical potentials these effects are
typically included in an average way, since they are naturally present in the
experimental data used to fit the potentials; some
potentials~\cite{Hellmann2008a} also use a semiempirical or approximate method
to include these effects.  But in order for a potential to systematically fit
the true potential energy surface it \emph{cannot} include quantum nuclear
effects at the level of the fitting, because the true Born-Oppenheimer potential
energy surface does not itself include these effects.  Thus, fitting methods that
include such an average contribution are not fitting the true potential energy
surface and are therefore incompatible with the current strategy.

The most common and practical technique for including quantum nuclear effects
(ZPVE and nuclear tunneling, but \emph{not} the nuclear exchange) in MD simulations is
via path integral molecular dynamics (PIMD), where the quantum system is
represented by $P$ replicas of the classical system, corresponding atoms being
joined across the replicas by harmonic springs in a ring-polymer
structure~\cite{Chandler1981,Habershon2013,Markland2018,Biermann1998}.  Recent
techniques, including improved stochastic
thermostats~\cite{Ceriotti2010a,Ceriotti2009a,Ceriotti2012} and
ring polymer contraction~\cite{Markland2008}, are making PIMD practical even for
large systems and more expensive potentials such as the ones employed in this
work.

Despite these new developments, \abinit\ liquid simulation remains a challenge.
The process of designing a machine learning potential for a new material,
especially for amorphous or liquid simulation, is still a laborious manual
process.  In this work we develop a methodology that will eventually serve as a
foundation for more systematic, perhaps even automated, development of
potentials for more complex molecular liquids.

\section{Model development methodology}
Fundamental to this methodology is a
strategy common to most successful potentials for molecular systems: The energy
of the system is decomposed into several terms that each represents a different
physical interaction.  From the point of view of a physics-based analytical
potential, this decomposition is useful because the different physical
interactions will typically have different functional forms, and it makes sense
to parameterize them separately.  From the point of view of a machine learning
potential, the main advantage of an energy decomposition scheme is that it
separates physical effects that take place at different length and energy scales
and prevents the larger effects from overwhelming the smaller ones; while the
smaller components might not be important in reproducing the \emph{total}
energy, other important observables (such as the density or the diffusivity)
might well weight these contributions much higher.  By controlling the accuracy
of the several components separately it is possible to achieve good accuracy on
any  property of interest.

In a molecular liquid such as methane, the primary separation in energy scales
is between the strong intramolecular (covalent) interactions and the weak
intermolecular (noncovalent) interactions.  These two types of interactions are
easy to separate and have characteristic energy scales that are orders of
magnitude apart.  The second separation we will employ here is motivated by the
length scales of the interactions, as machine learning potentials tend to work
best for fitting functions that vary on a
single length scale.  In methane, the dispersion (van der Waals) interaction
is very long-ranged, being still relevant at \atompair{C}{C} distances as large
as \SI{15}{\angstrom}, but the various repulsive interactions generated by
electron cloud overlap die out by \atompair{C}{C} distances of
\SI{5}{\angstrom}.
The long-range electrostatic energy is predicted (by most
empirical potentials) to be negligible for pure methane, which is neutral and
nonpolar, so it is not explicitly included.  The energy equation we
will use is therefore:
\begin{equation}
    E_\text{total} = E_\text{1b} + E_\text{repulsion,b1b} + E_\text{dispersion}
    \label{eq:energy-decomp}
\end{equation}
where the `1b' (one-body) energy is the covalent part and `b1b' signifies 
the intermolecular (beyond one-body) part.  The second term here is computed
from DFT beyond-one-body interactions, while the dispersion term is computed
separately.

Besides making it easier to fit the potential, another advantage of this
approach is that it allows us to capture some of the underlying physics of the
system.  Some recent analytical potentials take the approach of more closely
representing the underlying physics by extracting forcefield parameters from
fundamental physical quantities such as the electron density.  Models using this
approach include the Slater-ISA model of Van Vleet \etal~\cite{VanVleet2016} and
the biomolecular force field of Cole \etal~\cite{Cole2016}.  The IPML model of
Bereau et. al.~\cite{Bereau2018} goes one step further by using machine learning
to efficiently predict these properties across chemical compound space.  While
the physical interpretability of these models is appealing, it comes at the cost
of sacrificing a best-possible fit to the true quantum potential energy surface.
In the present potential, by capturing the simple, physically motivated parts
of the energy expression by simple analytical forms and fitting the complex,
nonanalytical parts as corrections on top of these, we use physics to guide our
description of the interaction while maintaining complete flexibility of the
functional form.

\begin{figure}[ht]
    \centering
    \includegraphics[scale=0.8]{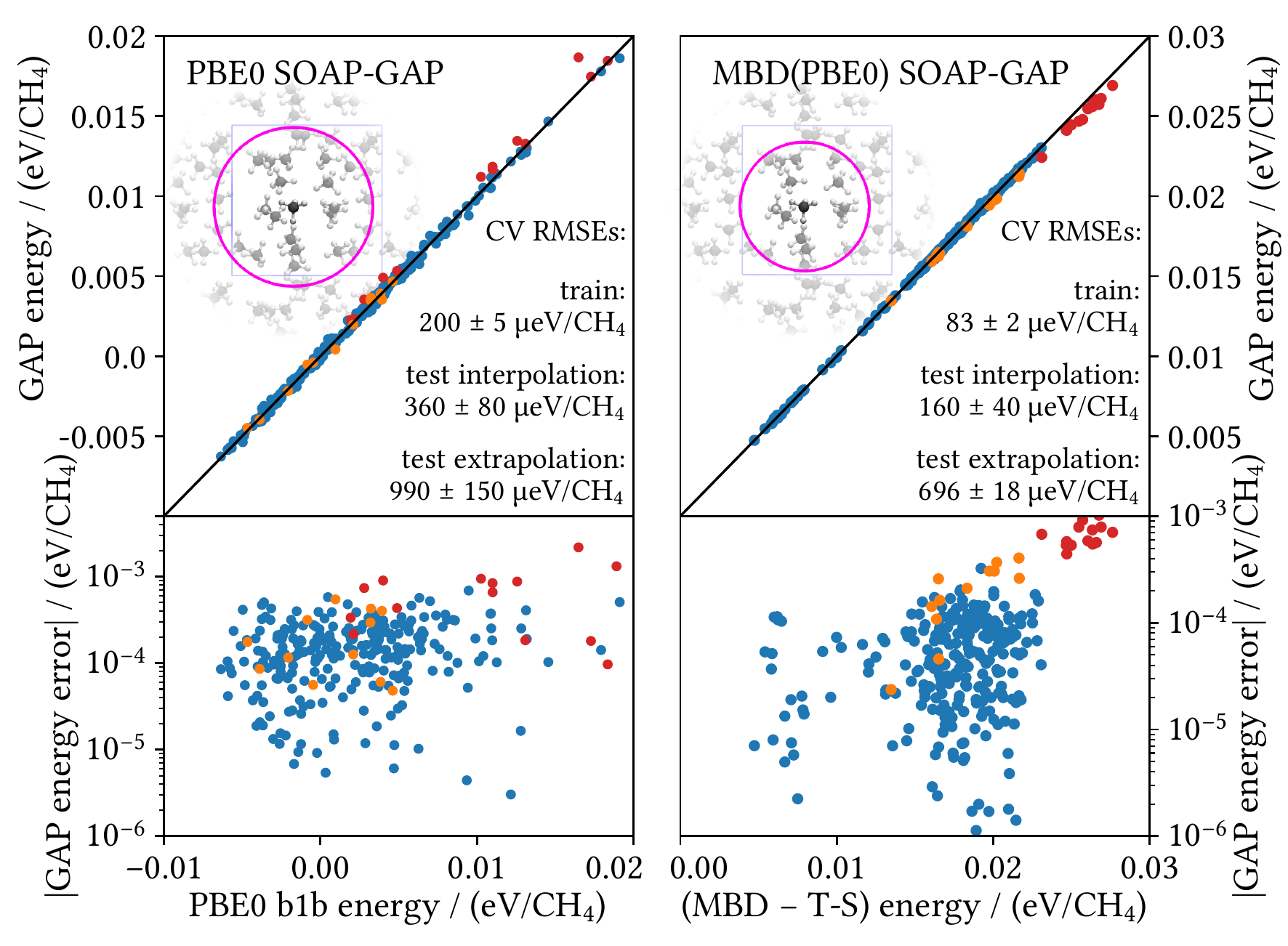}
    \caption{The PBE0 and MBD(PBE0) SOAP-GAP fits on 258 cell
        interaction (beyond one-body, `b1b') energies and (only for PBE0)
        corresponding forces.  Top: Correlation
        plots with the line $y=x$ of perfect correlation.  Bottom: Errors on a
        logarithmic scale.  The blue points represent the training set.  The
        orange points represent the interpolation test set and the red points
        represent the extrapolation test set (color online), neither of which
        was used in training the model.}
    \label{fig:gap-eval}
\end{figure}

\subsection{Machine learning model}
\begin{figure}[ht]
\centering
\includegraphics[scale=0.9]{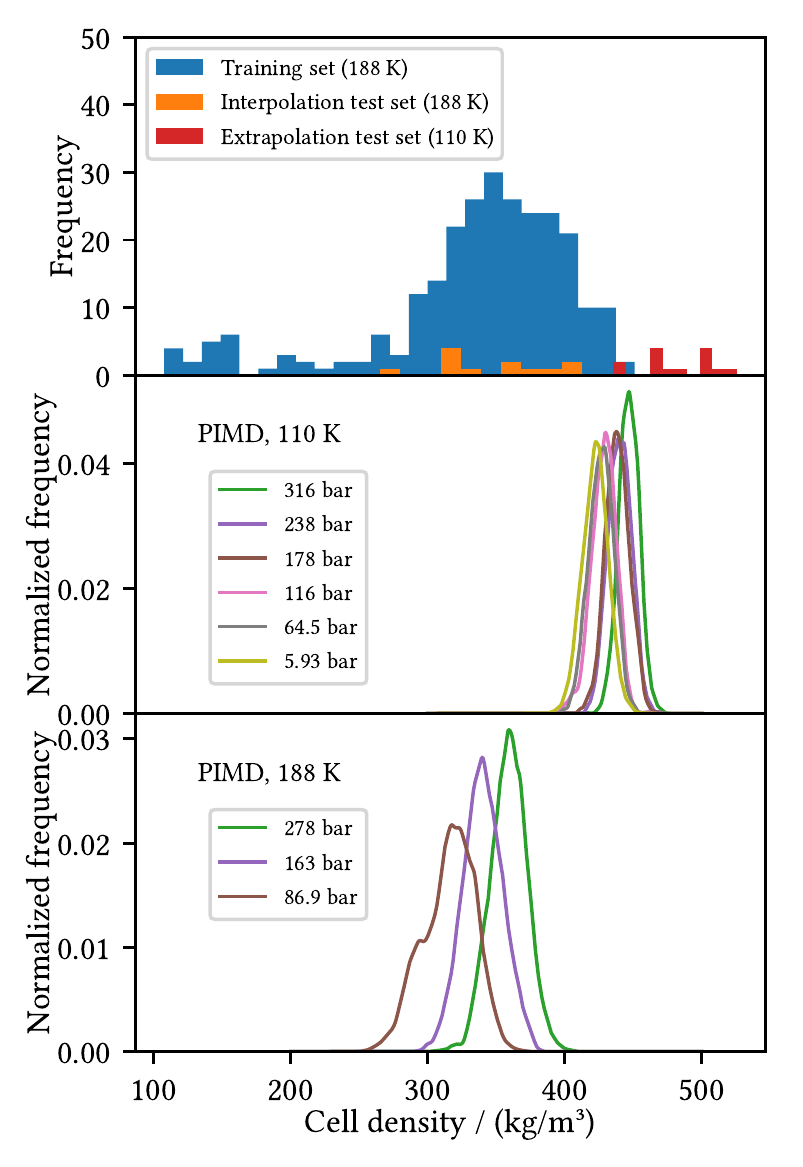}
\caption{Histograms over mass density of the cells in the training and two
         test sets, interpolation and extrapolation.  The
         distributions of densities encountered in the subsequent PIMD
         simulations with the (PBE0 SOAP)/COMPASS + T-S + MBD(PBE0) SOAP model
         (see below) are shown below for comparison.}
\label{fig:density-histograms}
\end{figure}

To fit the nonanalytical components of the potential for bulk methane --
the second part of Equation~\eqref{eq:energy-decomp} -- we use
the GAP method~\cite{gap-prl,Bartok2015a} with the SOAP kernel~\cite{soap-prb},
both developed and used by our group to fit complex, many-body potentials.  The
SOAP-GAP potentials were fit to DFT~\cite{DFT-HK,DFT-KS} b1b (beyond-one-body
interactions; the monomers were computed separately and subtracted from the
total cell) energies and forces, computed on samples of bulk methane taken from
MD trajectories under liquid conditions run using a
classical potential (OPLS/AMBER~\cite{Jorgensen1988,amber}) at a temperature of
\SI{188}{\kelvin} and five pressures ranging from \SIrange{0}{400}{\bar}, thus
covering the entire range of pressures encountered in the subsequent GAP MD
simulations.  The resulting training set consisted of a wide range of densities;
see Figure~\ref{fig:density-histograms}.  However, the typical densities
encountered during a simulation at \SI{110}{\kelvin} in the same pressure range
fall partly outside this range, exercising both the model's interpolation and
extrapolation capabilities.  To validate these capabilities, independent samples
were drawn from OPLS/AMBER simulations at both temperatures, with several
samples taken from each of the state points where classical results are shown in
Figure~\ref{fig:isotherms} below.  The histogram of the densities of these test
sets is also shown in Figure~\ref{fig:density-histograms}.  Based on the
position of these distributions relative to the test set, the 12 test samples taken
at \SI{188}{\kelvin} were labeled the `interpolation' test set and the 14 samples
from \SI{110}{\kelvin} were labeled the `extrapolation' test set.

The DFT calculations on all cells were done using CASTEP~\cite{CASTEP}.  Two
functionals were used, PBE~\cite{Perdew1996} and the hybrid GGA functional
PBE0~\cite{Adamo1999}.  The GAP fits were done using the SOAP
descriptor~\cite{soap-prb}, resulting in two models called \mbox{`PBE SOAP-GAP'}
and \mbox{`PBE0 SOAP-GAP'}.  The performance of the \mbox{PBE0 SOAP-GAP} is assessed in
Figure~\ref{fig:gap-eval}, which indicates good reproduction of both energies
and forces on the training set.  Since GAP is a statistical learning method, 
this is usually a good measure of how the method will perform on similar geometries.
The interpolation performance indicates some degree of overfitting, though,
while the extrapolation performance is significantly worse -- but the model
still achieves an error of less than \SI{1}{\milli\electronvolt} per molecule on
systems that were never included in the fit.  The variability of this error
measure was assessed with a cross-validation (CV) procedure: Ten disjoint sets
of twelve points each were selected from the
training data, and each in turn substituted with the interpolation test set to
train ten additional GAP models.  The numbers reported in Figure~\ref{fig:gap-eval}
are obtained as the mean and standard deviation of the errors across this set of
eleven GAPs, with the withheld points standing in for the interpolation test set
in each validation GAP.
The errors on the forces show the same pattern: The training set error is
\SI{6.56(3)}{\milli\electronvolt\per\angstrom}, the interpolation test set error is
\SI{6.8(6)}{\milli\electronvolt\per\angstrom}, and the extrapolation test set
error is \SI{8.71(5)}{\milli\electronvolt\per\angstrom}.  Plots of the forces
for the similar \mbox{PBE SOAP-GAP}, along with its energy and force errors, can be found in the
supporting information.

\subsection{Dispersion model}
The dispersion component, the third term in Equation~\eqref{eq:energy-decomp}, was
accounted for using two levels of theory.  The
first was the pairwise correction of Tkatchenko and
Scheffler~\cite{Tkatchenko2009}.  This method uses relative atomic volumes from a
Hirshfeld partitioning~\cite{hirshfeld-charges} of the electron density, an idea
introduced by Becke and Johnson~\cite{Becke2006}, and relates them to free-atom
dispersion coefficients (those computed by Chu and Dalgarno~\cite{Chu2004} were
used here).  Recomputing the Hirshfeld volumes for each step of an MD simulation
would be impractically expensive, as that would require a new DFT calculation at
each step.  Instead, the first level of theory only uses the per-element average
of the relative Hirshfeld volumes across the sample of DFT cells.  The
dispersion correction can then be applied as an analytical pair potential whose
form and parameters are fixed throughout the simulation, a scheme hereafter
termed simply `T-S'.

The second level of theory is the MBD, or many-body dispersion,
method~\cite{Tkatchenko2012,Ambrosetti2014}.  Despite the greater complexity of
the MBD approach, it can still be viewed as a correction on top of the pairwise
Tkatchenko-Scheffler interaction.  Thus, another SOAP-GAP was fit to the difference
between the MBD energies only and the (fixed) T-S term as the baseline, once each for
PBE and PBE0 Hirshfeld volumes.  This model, termed `MBD(PBE) SOAP-GAP' (and the
corresponding `MBD(PBE0) SOAP-GAP'), accounts for relatively short-ranged
many-body effects.  The
dispersion energy term from Equation~\eqref{eq:energy-decomp} therefore becomes:
\begin{equation}
    E_\text{dispersion} = E_\text{T-S} + E_\text{MBD SOAP-GAP}.
\end{equation}
The MBD SOAP-GAP also implicitly accounts for the variability of the
Hirshfeld volumes that was neglected in the fixed T-S model: The SOAP
descriptor is sensitive to the intramolecular and short-range geometrical
factors that (presumably) also account for the variability of these volumes.  The
MBD(PBE0) fit is likewise assessed in Figure~\ref{fig:gap-eval}, showing that
both its interpolation and extrapolation performance is similar to that of the
PBE0 SOAP-GAP.

Finally, a complete model for liquid methane must also include an intramolecular
component (the first term in Equation~\eqref{eq:energy-decomp}).  Two empirical
potentials are considered for this purpose: AMBER~\cite{amber} includes only
harmonic bond and angle terms, while COMPASS~\cite{Sun1998} includes
higher-order anharmonic and cross-coupling terms.  Both models were tested in
order to help measure the influence of such effects (anharmonic and
cross-coupling) on the predicted properties, especially with the inclusion of
quantum nuclear effects.

\section{Results}
The first test of the accuracy and applicability of any potential for liquids
is how well it reproduces the experimental equation of state.  While most
empirical potentials (for example OPLS~\cite{Jorgensen1988}) are fit to
reproduce experimental thermodynamic data, the fitting conditions are often
only a single state point per material, usually standard temperature and
pressure.  Some potentials, like TraPPE~\cite{Martin1998}, are fit to reproduce
thermodynamic data across a wide range of state points, in this case by fitting
coexistence curves.  Therefore, a wide range of temperature and pressure
conditions were chosen to test the accuracy of the potentials considered.  Two
isotherms were chosen where experimental data was available
(from Goodwin and Prydz~\cite{Goodwin1972}): At \SI{110}{\kelvin}, density
measurements were available at
\SIlist{5.93;64.5;116;179;238;316}{\bar}\cite{presstrunc}.  At
\SI{188}{\kelvin}, density measurements were available at
\SIlist{86.9;163;278}{\bar}\cite{presstrunc}.

\begin{figure}
\centering
\includegraphics[scale=0.55]{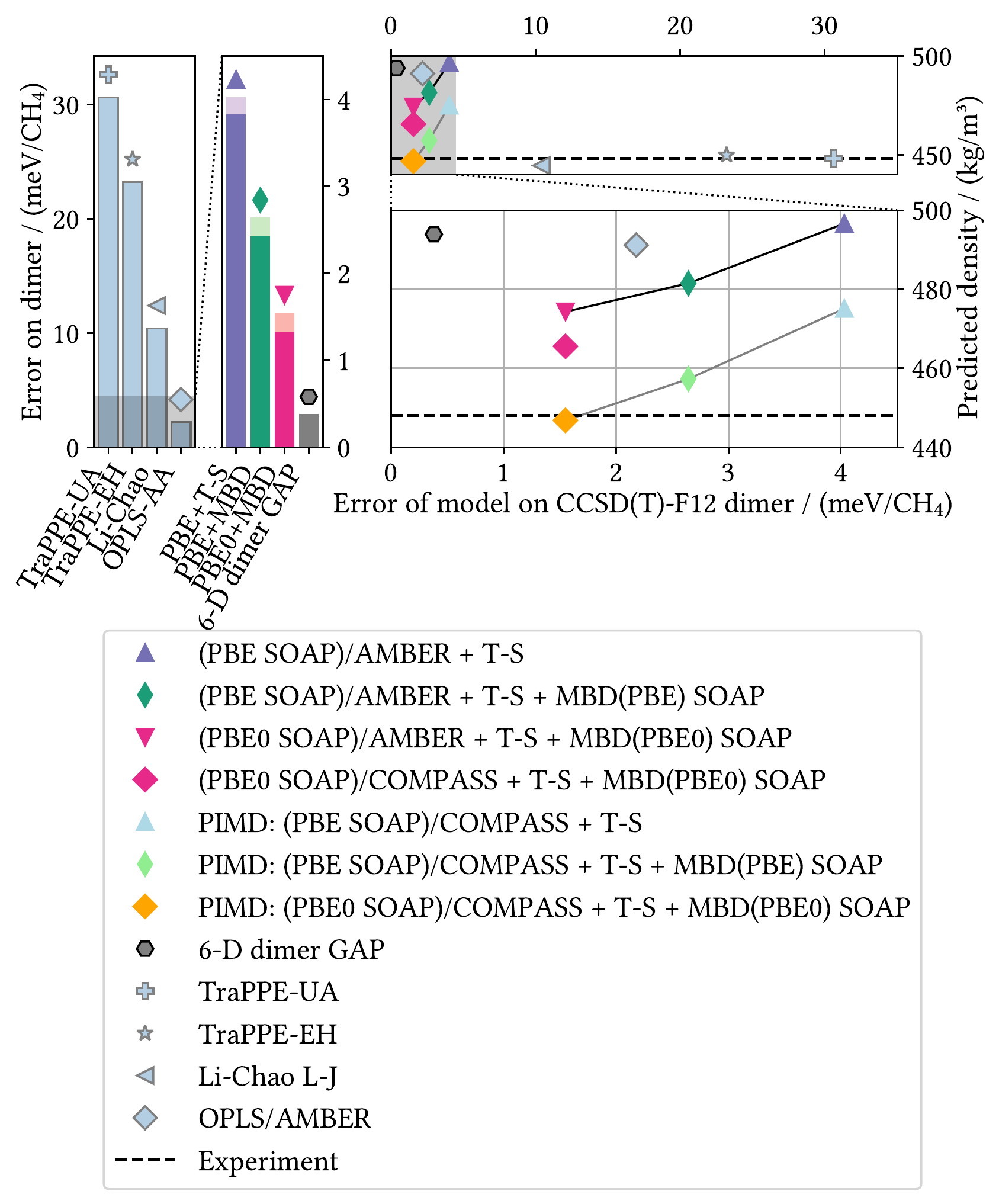}
\caption{Systematic convergence of the density predictions of the bulk
    SOAP-GAPs as the dispersion model is improved; single state point at
    \SI{110}{\kelvin} and \SI{316}{\bar}: Convergence of the classical
    simulations is shown by the black connecting line, PIMD simulations with the
    grey line. The uncertainties on the density are smaller than the sizes of
    the symbols.  The ``dimer error'' figure is computed against CCSD(T)-F12 on
    the sample of dimers used to train the 6-D dimer GAP.  In the right-hand bar
    plot, solid bars represent the systematic errors due to the underlying
    quantum model and the pastel bars on top represent the statistical errors
    introduced by the GAP fit.  In the left-hand bar plot, the bars represent
    the (systematic) error of the analytical model against the same
    coupled-cluster reference.}
\label{fig:dimer-error-convergence}
\end{figure}

Four GAPs were chosen for testing: The `PBE SOAP-GAP' model with both fixed T-S
(`\mbox{+ T-S}')
and MBD (`\mbox{+ T-S} \mbox{+ MBD(PBE) SOAP}') dispersion, the
`\mbox{PBE0 SOAP-GAP} \mbox{+ T-S} \mbox{+ MBD(PBE0) SOAP-GAP}',
and the 6-D dimer GAP described above and shown in Figure~\ref{fig:dimer-anisotropy}.
The dimer GAP and all of the SOAP-GAP models were first tested at the state
point \SI{110}{\kelvin} and \SI{316}{\bar} using a ``smart sampling''
colored-noise thermostat for efficient equilibration~\cite{Ceriotti2010}.  The
convergence of the results towards the experimental density
is illustrated in Figure~\ref{fig:dimer-error-convergence}; for brevity, all the
`SOAP-GAP' models are labeled simply with `SOAP'.  A selection of empirical
potentials is also shown: TraPPE-UA~\cite{Martin1998},
TraPPE-EH~\cite{Chen1999}, OPLS/AMBER~\cite{Jorgensen1988,amber} (a combination
of OPLS-AA with the AMBER intramolecular parameters), and the
Li-Chao \lj\ fit~\cite{Li2016}.

The density predictions are shown
against the error of the \emph{underlying} quantum model computed on a
sample of dimers (the same one used to fit the 6-D dimer GAP), with
\mbox{CCSD(T)-F12}
taken as the reference.  In the case of the GAPs, the statistical
uncertainty introduced by the fits is shown and added to the systematic
uncertainty already given by the quantum model.  In the case of the empirical
model, the error is taken to be entirely systematic.

Evidently, the predictions for the density at both state points improve as the dispersion model
is made more accurate as measured on the methane dimer, an improvement that is
reflected in the dimer error measure.  Adding the MBD SOAP-GAP lowers
the density by \SI{15}{\kgpcm}, improving the prediction by \SI{3.4}{\percent}
with respect to experiment; the short-range improvement offered by
switching to PBE0 gives a further \SI{7.2}{\kgpcm} (\SI{1.6}{\percent})
improvement.  Given the large influence
many-body effects appear to have on the density, it is perhaps not
surprising that the general dimer GAP gives such a bad prediction,
\SI{45.8}{\kgpcm} or \SI{10.2}{\percent} higher than experiment.
While the figure
indicates that there are still effects not included by the dimer measure of
accuracy -- especially the intramolecular potential and many-body (beyond dimer)
effects~-- it
still shows a general trend of improvement of the potential's predictions as it
more accurately represents the underlying potential energy surface.
Crucially, this is a trait not shared by empirical potentials -- TraPPE,
OPLS/AMBER and the Li-Chao L-J -- which show the opposite behavior.  The models
with the best density prediction are also the ones with the worst performance on
the methane dimer, indicating that they must achieve their accuracy by a large
cancellation of errors.  (It should also be noted that the accuracy of the
Li-Chao L-J model is without explicit quantum nuclear effects, even though the
potential was ostensibly fitted to the Born-Oppenheimer potential energy surface
of the dimers, thus further muddling the picture for empirical potentials). 

\begin{figure}[t]
\includegraphics{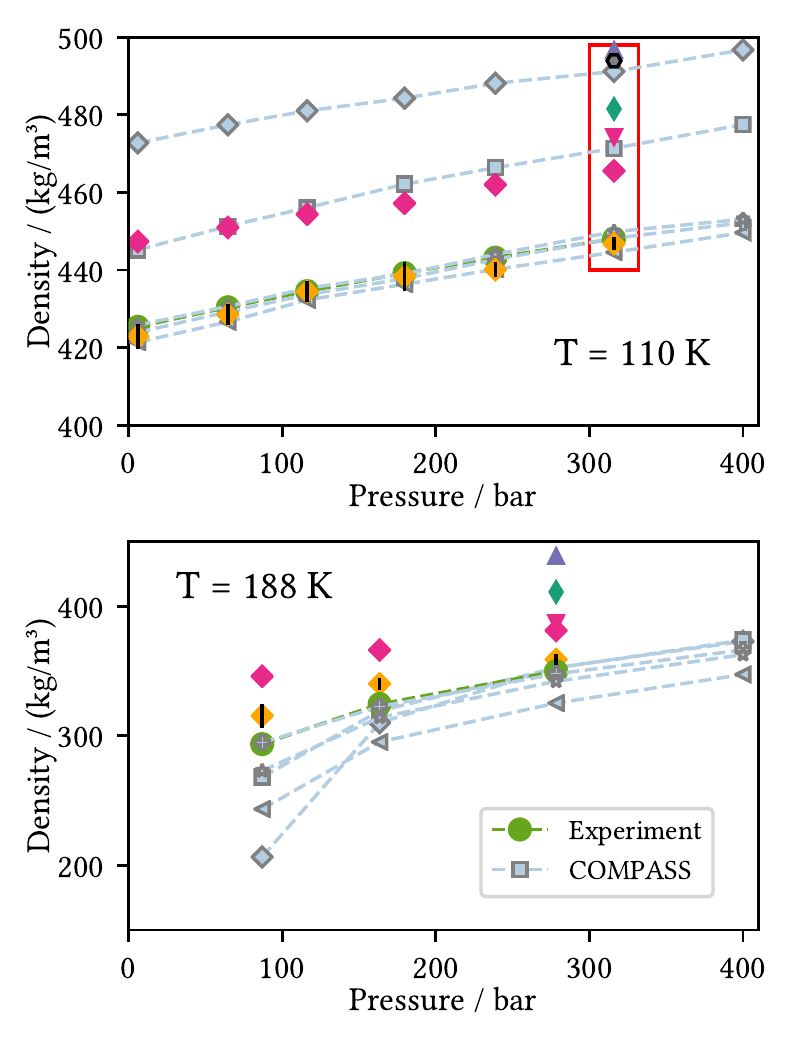}
\caption{Equation of state at two temperatures, \SI{110}{\kelvin} and
\SI{188}{K}, as predicted by various atomistic models.  The bulk SOAP-GAPs with
different dispersion models are shown, as is the 6-D dimer GAP.  All-atom
empirical models are shown in gray. 
Experimental data from Goodwin and Prydz~\cite{Goodwin1972}.  The small black
lines are error bars on the PIMD simulations computed using the blocking method
described in the supporting information.  Refer to the legend of
Figure~\ref{fig:dimer-error-convergence} for symbols previously defined.}
\label{fig:isotherms}
\end{figure}

The quantum nuclear effect was assessed in an explicit way,
using a PIMD simulation using the PIGLET
thermostat~\cite{Ceriotti2009a,Ceriotti2012}.  With this effect included, the best
model (`PBE0 SOAP + T-S + MBD(PBE0) SOAP') delivers a prediction within
\SI{0.3}{\percent} (nearly within simulation uncertainty) of the
experimental number.  This decrease in density is of the same order of magnitude as that
reported in~\citet{Pereyaslavets2018}, though with this potential the effect is
smaller -- \SI{4.2}{\percent} instead of \SI{9}{\percent}.
Figure~\ref{fig:isotherms} shows that the size of the effect is roughly the same
across the \SI{110}{\kelvin} isotherm, so even at the \SI{112}{\kelvin},
\SI{1}{\bar} state point used in that study we would expect to
see a somewhat smaller effect.
The decrease also intuitively contrasts with the finding that,
in the gas phase of methane, zero-point vibrational contributions
\emph{increase} the molecular $C^6$ (first pairwise dispersion) coefficient and
hence the strength of the intermolecular
attraction~\cite{Hellmann2008a,Russell1995,Bishop1998}.
Evidently, this effect is overpowered in the condensed phase by an increase in
repulsion leading to a decrease in the density; a likely candidate is the
increase in molecular size due to the same zero-point vibrational energy
lengthening the bonds~\cite{Balog2000}.  The exact mechanism is still
unclear, though, and the \abinit\ quality potentials presented here provide the
means necessary for further study of this effect.

The performance of the models across both of the experimental isotherms is shown
in Figure~\ref{fig:isotherms}.  For comparison, a selection of analytical
potentials was tested at all the state points at \SI{110}{\kelvin} and
\SI{188}{\kelvin} with experimental data, plus an additional point at
\SI{400}{\bar} for each isotherm to show the high-pressure trend.  In addition
to the potentials shown in Figure~\ref{fig:dimer-error-convergence}, the figure
also shows the \ljbaseline\ and COMPASS~\cite{Sun1998}.  Note in particular
that the empirical all-atom potentials all shift with respect to experiment
between the two isotherms.  Most models, the GAPs included, have more trouble
reproducing the density at the \SI{188}{\kelvin} isotherm, perhaps because of
the proximity of the lowest-pressure point to the critical point
(\SI{190.58}{\kelvin} and \SI{46.04}{\bar}~\cite{Teja1990}).  Only the
united-atom model TraPPE-UA maintains
accuracy across the whole space of conditions covered, with the
explicit-hydrogen description TraPPE-EH closely following in consistency.  The
series of SOAP-GAP potentials delivers predictions of increasing accuracy, in
correlation with the accuracy on the dimer.
Despite the relatively large statistical fluctuations in the PIMD SOAP-GAP
density predictions, the model is still more consistently accurate (comparing
across both isotherms) than any other model fit to the quantum PES, especially
with the explicit inclusion of quantum nuclear effects.  It thus appears
essential to include quantum nuclear effects in order to make accurate
predictions with a potential fitted to the Born-Oppenheimer quantum potential
energy surface.  Other potentials that achieve agreement with experiment
without explicit treatment of these effects must be incorporating them into the
potential energy surface itself, which is at odds with our stated goal of
achieving the agreement with experiment in an {\em ab initio} manner by best
fitting the potential energy surface.

In summary, while TraPPE potentials
obtain their accuracy by fitting to experimental data across wide ranges of
temperature and pressure, the SOAP-GAP potentials obtain their accuracy by
fitting to the underlying quantum mechanical description of matter and
systematically converge to within \SI{0.5}{\percent} of the experimental value
as their description is improved (while TraPPE-EH, while
being in some sense a more accurate model than TraPPE-UA by describing the
anisotropy of the methane molecule, delivers worse predictions especially at the
\SI{188}{\kelvin} isotherm).  Additionally, even the current best SOAP-GAP model
still has several routes of potential improvement that would not be open to a
fixed-form analytical potential, such as changing the intramolecular model for a
more accurate, fitted one or improving the dimer description to the
coupled-cluster dimer GAP level (which can be done using existing techniques,
e.g. by adding a further two-body correction to the SOAP-GAP
model~\cite{Bartok2013,Gillan2013}).

\section{Discussion}
The fitting and testing of the SOAP-GAP and dimer potentials for
liquid methane reveal three key findings for the description of molecular
liquids: First, many-body effects -- not only within the dimer, but also
beyond-dimer effects -- are essential, especially in the short range,
for obtaining an accurate description of the bulk density.  Second, an explicit
description of quantum nuclear effects is equally important, especially at the
temperatures and pressures considered here.  Third, systematic measures of the
accuracy of the potential (such as the dimer error measure presented here) are a
good guide to improving systematically fitted potentials toward convergence with
the experimental results, a goal which the \mbox{(PBE0 SOAP-GAP)} \mbox{+ T-S}
\mbox{+ MBD(PBE0) SOAP-GAP} model presented here comes close to achieving.

The methodology presented here is a new, physics-based, systematic path toward
creating exceptionally accurate potentials for molecular liquids.  The
methodology is expected to generalize in a straightforward way to longer
hydrocarbons, where the more difficult part then becomes the description of
intramolecular interactions.  Furthermore, the ideas presented here could be
extended to other types of long-range interactions, such as electrostatics and
induction, in order to extend accurate machine learning potentials to a wider
variety of molecular liquids.

\section{\label{sec:methods}Computational Methods}

\subsection{\label{sub:gaussian-processes}Gaussian processes}

The GAP machine learning method used to fit the potential energy is based on
Gaussian process regression and is part of the family of kernel learning
methods~\cite{gap-prl,Bartok2015a}.  Such methods perform linear fits in a
transformed data space: The nonlinearity of the function is now captured in a
kernel function, also called a similarity or covariance function, which usually
measures the similarity between two local atomic environments (although they
can also be designed to capture long-range and global properties).

Formally, the potential energy suface is represented as a Gaussian
process~\cite{itila,Rasmussen2006}.  The covariance matrix of this process is
formulated to use the information provided by quantum calculations, i.e. total
energies and derivatives, in a natural way through linear operations on the
kernel.  This allows the Gaussian process to provide a smooth approximation of
the potential energy surface, as sampled by the quantum data points, in the high-dimensional space of atomic or molecular
environments using just a linear combination of kernels; for example, the local
energy of an atom $i$ is given by:
\begin{equation}
    \varepsilon_i = \sum_j \alpha_j k(\bvec{d}_j, \bvec{d}_i)
\end{equation}
where the ${\bvec{d}}$ are descriptors of local atomic environments, $k$
designates the covariance or kernel function, and the weights ${\alpha}$ are
determined by a regularized least-squares linear fit to the quantum mechanical
training data (in this view, the predictions of Gaussian processes are the same
as those given by kernel ridge regression (KRR) with a radial
basis)~\cite{Bartok2015a}.  In GAP, the sum runs over a subset of
\emph{representative} configurations in the training set, allowing the fitting
to scale \emph{linearly} with the number of input data points.

The most successful kernel function for condensed-phase GAP has been the SOAP
kernel~\cite{soap-prb}, which takes the similarity between local atomic
environments.  The environment of atom $i$ is represented by a neighbor density
$\rho_i(\bvec{r})$, defined as a sum of Gaussians placed on each neighboring
atom, multiplied by a spherical cutoff function which smoothly takes the density
to zero outside some cutoff radius.  The kernel between two environments is
defined as the integral over all possible mutual rotations of the square of the
overlap between the two neighbor densities, thus making the kernel obey the same
symmetries as the local energy: Invariance to translations (environments are
atom-centred), permutations (from summing like atoms in the neighbor density),
and rotations (from the rotational integration).

In practice, the integration over rotations can be done analytically by expanding each
neighbor density in spherical harmonics and radial basis functions:
\[
    \rho_i(\bvec{r}) = \sum_{nlm} c_{nlm}^{(i)} g_n(r) Y_{lm}(\hat{\bvec{r}}),
\]

computing the power spectrum elements

\begin{equation*}
    p_{nn'l}^{(i)} = \frac{1}{\sqrt{2l + 1}} \sum_m c_{nlm}^{(i)}
    (c_{n'lm}^{(i)})^\dag,
\end{equation*}

and summing in order to obtain the covariance function:

\begin{equation}
    k_0(\rho_i, \rho_j) = \sum_{nn'l}p_{nn'l}^{(i)}p_{nn'l}^{(j)}
\end{equation}

which is then normalized to obtain a proper kernel and optionally raised to some
power $\zeta > 1$ to increase the sensitivity to changes in the local
environment\cite{soap-prb,Bartok2015a}.

Note here that the local environment is effectively represented by a set of
numbers $\bvec{p}^{(i)}$, which can be interpreted as a ``descriptor'' or even
``feature vector'' of the environment.  Many other kernels are formulated in
terms of such descriptors, such as the 6-D dimer kernel described in the
supporting information.

The GAP models used in this study were all fit and evaluated using the
libAtoms/QUIP package~\cite{libAtoms}.  The GAP code can be
downloaded at \url{http://www.libatoms.org/gap/gap_download.html}, with a
precompiled version available through Docker at
\url{https://hub.docker.com/r/libatomsquip/quip/}.

\subsection{\label{sub:md-methods}MD simulations}

The MD simulations were run using QUIP~\cite{libAtoms} and
i-PI~\cite{Ceriotti2014} via LAMMPS~\cite{lammps,lammps11Aug17}.  The former used
the adaptive Langevin thermostat of Jones and Leimkuhler~\cite{Jones2011} and
a Hoover-Langevin barostat~\cite{Quigley2004} while the latter used a thermostat based on the generalized Langevin equation (GLE, otherwise known as colored-noise thermostats), namely the
``smart sampling'' method of Ceriotti, Bussi, and
Parrinello~\cite{Ceriotti2010}, for the classical simulations and
PIGLET~\cite{Ceriotti2009a,Ceriotti2012} for the PIMD simulations.
The initial configurations for all simulations were generated using
Packmol~\cite{Martinez2009}.

The analytical potentials were run in LAMMPS~\cite{lammps} with a Langevin
thermostat~\cite{Bruenger1984} and a Nosé-Hoover
barostat~\cite{Nose1984,Hoover1985,Parrinello1981,Shinoda2004,Tuckerman2006}
with the MTK correction~\cite{Martyna1994}.  For potentials with a Coulomb component
(OPLS/AMBER and COMPASS), the contributions beyond the cutoff were calculated
with the particle-particle particle-mesh (PPPM) method~\cite{Hockney1988}.

\subsection{\label{sub:dimer-fits}Dimer fits}

The coupled-cluster CCSD(T) energies of the methane dimer were computed in a
similar way as described in Gillan et. al.~\cite{Gillan2013} (explained in more
detail in the supporting information), up to the level of
CCSD(T)-F12~\cite{Adler2007,Knizia2009,Kong2012}.  The energies were corrected
for basis-set superposition error (BSSE) using the Boys-Bernardi counterpoise
procedure~\cite{Boys1970}.  Calculations were done using the \mbox{MOLPRO} suite
of programs~\cite{MOLPRO,MOLPRO_WIREs,molpro-idirect,molpro-alaska}.  The
Atomic Simulation Enviroment (ASE)~\cite{ase} was used to generate and
manipulate geometries.  For the dimer error numbers used in
Figure~\ref{fig:dimer-error-convergence}, energies (PBE and PBE0) were computed
with \textsc{Psi4}~\cite{psi4} and the Hirshfeld
partitioning~\cite{hirshfeld-charges} was done using
HORTON~\cite{horton,becke1988_multicenter,becke1988_poisson,lebedev1999}.

The geometries for the randomly chosen orientations were directly sampled from a
liquid MD simulation (details in the supplementary information).  Ten
orientations were sampled and each used to produce a binding curve with
regularly spaced dimer separations.

Finally, all the graphics in this paper were made using
Matplotlib~\cite{Hunter:2007}; the analysis was done within the Jupyter
interactive computing enviroment with the IPython kernel~\cite{ipython}.

\begin{acknowledgments}
M.V. acknowledges Shell Global Solutions International B.V. for funding, as
well as support from the EPSRC Centre for Doctoral Training in Computational
Methods for Materials Science (under grant number EP/L015552/1).  This
work used the ARCHER UK National Supercomputing Service
(\url{http://www.archer.ac.uk}) under the UCKP Consortium, \mbox{EPSRC} grant
number EP/P022596/1.  We gratefully acknowledge the assistance of Venkat
Kapil in preparing the GLE and PIMD simulations.
\end{acknowledgments}

\section*{Supporting information}

Dimer fit details, parameters for the DFT and quantum chemistry
calculations, GAP fitting command lines, and MD simulation trajectories and
parameters are given in the supporting information.

The GAP definition files and parameter files required to reproduce the MD
simulations in this work are available online at
\url{https://doi.org/10.17863/CAM.26364}.

\bibliography{refs}

\end{document}


\title{Supporting information for:\\
       ``Equation of state of fluid methane from first principles\\
   with machine learning potentials''}

\author{Max Veit}
\email{max.veit@epfl.ch}
\altaffiliation{
Current address:
Laboratory of Computational Science and Modeling,
Ecole Polytechnique Fédérale de Lausanne,
1015 Lausanne, Switzerland
}
\affiliation{
Engineering Laboratory\\
University of Cambridge\\
Trumpington Street\\
Cambridge, CB2 1PZ\\
United Kingdom
}

\author{Sandeep Kumar Jain}
\author{Satyanarayana Bonakala}
\author{Indranil Rudra}
\affiliation{
Shell India Markets Pvt. Ltd.\\
Bengaluru 562149\\
Karnataka, India
}

\author{Detlef Hohl}
\affiliation{
Shell Global Solutions International BV\\
Grasweg 31\\
1031 HW Amsterdam\\
The Netherlands
}

\author{Gábor Csányi}%
\affiliation{
Engineering Laboratory\\
University of Cambridge\\
Trumpington Street\\
Cambridge, CB2 1PZ\\
United Kingdom
}

\date{\today}

\maketitle

\section{\label{sec:dimer-gap}Dimer energies}

The binding curves of the methane dimer shown in Figure~1 of the main paper were
computed in a similar way as described in Gillan et. al.~\cite{Gillan2013}: the
Hartree-Fock (HF) energy was computed at the largest basis, the Dunning
correlation-consistent basis set~\cite{Dunning1989,Kendall1992} aug-cc-pV5Z
(hereafter called AV5Z).  The energy difference between MP2 and HF was computed
using the smaller AVQZ basis.  Finally the difference between CCSD(T) (with
explicitly correlated basis functions, called
CCSD(T)-F12a~\cite{Adler2007,*Knizia2009,Kong2012}) and MP2 was computed using
the AVTZ basis.  The corrections were successively added to the base HF energy
to obtain energies at each of the HF, MP2, and CCSD(T)-F12 levels, and
additionally forces at the HF and MP2 levels.  Finally, all of the energies were
corrected for basis-set superposition error (BSSE) using the Boys-Bernardi
counterpoise procedure~\cite{Boys1970}.  Calculations were done using the
\mbox{MOLPRO} suite of programs~\cite{MOLPRO,*MOLPRO_WIREs,*molpro-idirect,
*molpro-alaska}.

The geometries for the symmetric orientations were generated using the Atomic
Simulation Environment (ASE)~\cite{ase} starting with monomers that had been
optimized at the MP2/AVQZ level, resulting in a \atompair{C}{H} bond length of
\SI{1.085}{\angstrom}.  The first two configurations correspond
exactly with configurations used in Chao \etal~\cite{Chao2009} and Hellmann
\etal~\cite{Hellmann2008a}; the figure gives the labels each author assigned to
these configurations.  The other three are similar, though not exactly the same,
as the corresponding labeled configurations.

A first dimer model was obtained in a similar way as in Li and
Chao~\cite{Li2016}, by fitting a pairwise \lj\ to the energies of the symmetric
orientations shown in the main paper.  The model was a standard 12-6 \lj\
between all atom pairs; the six coefficients for the different pair types were
all optimized by a least-squares fit.  The optimization produced \atompair{C}{H}
and \atompair{H}{H} potentials that were nearly purely repulsive, so the form
$\phi(r) = A r^{-12}$ was adopted for these instead.  The \atompair{C}{C}
potential has the standard \lj\ form: $\phi(r) = -4\epsilon ((r / \sigma)^{-6} -
(r / \sigma)^{-12})$.  The parameters of this model are given in
Table~\ref{tab:ljrep-params}.

\begin{table}[ht]
\begin{tabular}{c c S}
\toprule
Parameter & Pair type & \text{Value} \\
\colrule
$\sigma$ & C-C & \SI{3.52608}{\angstrom} \\
$\epsilon$ & C-C & \SI{0.00135}{\electronvolt} \\
$A$ & C-H & \SI{517.030}{\electronvolt} \\
$A$ & H-H & \SI{23.4878}{\electronvolt} \\
\botrule
\end{tabular}
\caption{Parameters for the optimized pairwise \lj\ model}
\label{tab:ljrep-params}
\end{table}

The fitting dataset used dimer distances from \SIrange{3.5}{9.5}{\angstrom} in
steps of \SI{0.5}{\angstrom}, with additional points at \SI{3.25}{\angstrom} and
\SI{3.75}{\angstrom}.  Energies larger than \SI{+0.02}{\electronvolt} were not
used in the fit.

This model was then taken as the \emph{baseline}, and further fits were done on
the difference between this \ljbaseline\ and the full energies in order to
improve upon this model.  For the new fits, a more thorough sample of the dimer
configuration space was needed, so a random sample of dimers was taken from a
liquid MD simulation using 200 rigid methane molecules with the monomer geometry
optimized at the composite CCSD(T)/AVTZ level described below (but without the
F12 correction), giving a
\atompair{C}{H} bond length of \SI{1.088}{\angstrom}, and the
intermolecular interactions computed using the OPLS-AA force
field~\cite{Jorgensen1996}; the simulation was run using the LAMMPS molecular
dynamics package~\cite{lammps,lammps5Oct15} at \SI{188}{\kelvin} and
\SI{400}{\bar} using a Langevin thermostat~\cite{Bruenger1984} and a Nosé-Hoover
barostat\cite{Nose1984,Hoover1985,Parrinello1981,Shinoda2004,Tuckerman2006}.
(The MD simulations used to generate the random orientations for the binding
curves in the main text were done the same way, except the monomers were fixed
with the OPLS-AA \atompair{C}{H} bond length of \SI{1.09}{\angstrom}.)
The dimers were sampled with a \atompair{C}{C} distance distribution from
\SIrange{3}{10}{\angstrom}, strongly favoring the short range of
\SIrange{3}{5.5}{\angstrom} and further enriched between
\SIrange{3}{3.5}{\angstrom}.  A section of this distribution is pictured with
the dimer binding curves in the main text; the full sample contains 2418 dimers.
The interaction energies of the dimers in this dataset were computed using the
same procedure as described for the fixed-orientation samples.

\begin{figure}
    \includegraphics[scale=0.9]{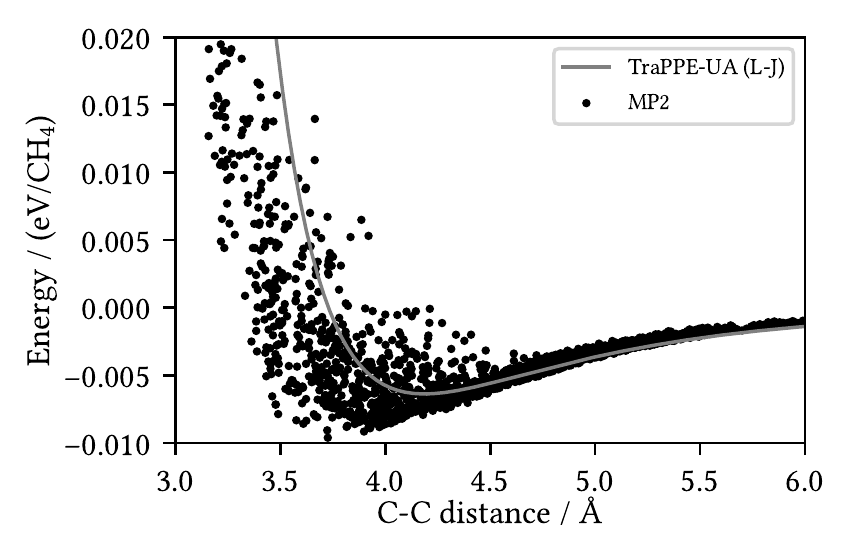}
    \includegraphics[scale=0.9]{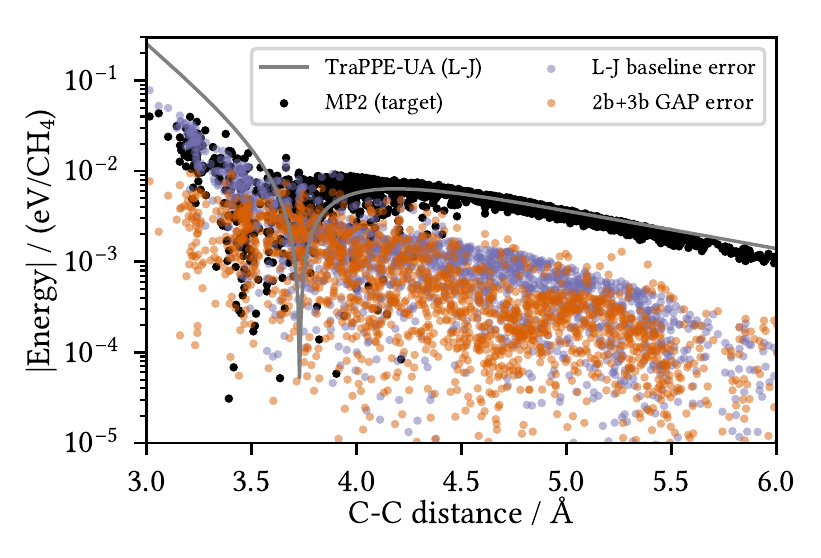}\\
    \hspace{0.18cm}
    \includegraphics[scale=0.9]{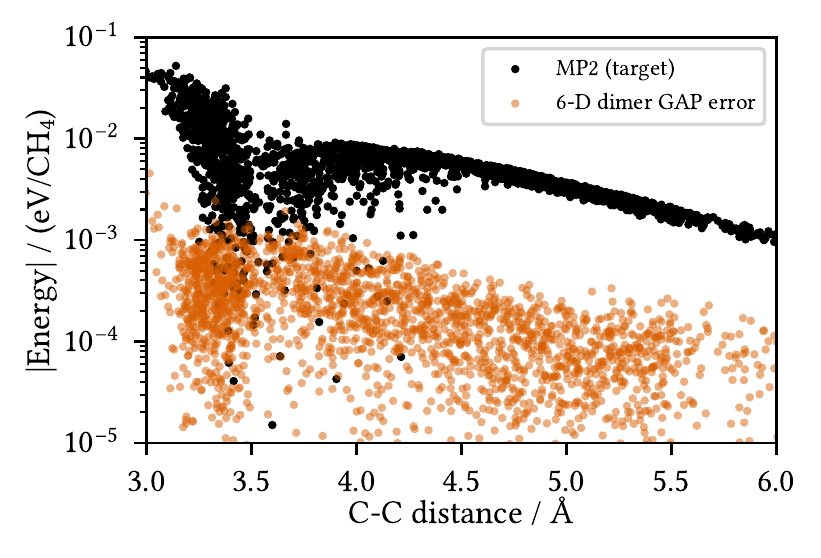}
    \includegraphics[scale=0.9]{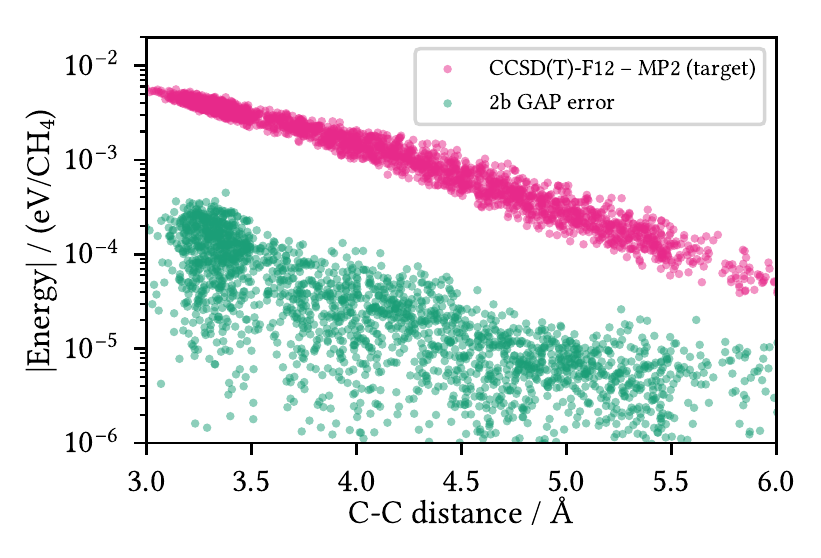}
\caption{Errors of successive GAP models fit to MP2 data, shown
as a function of C-C dimer separation.  The baseline is a pairwise L-J model
fitted to the coupled-cluster data from the symmetric orientations.
The first fit uses two-body and three-body descriptors, the second uses the
6-D dimer descriptor, and the final correction to the coupled-cluster
level is a simple two-body (pairwise) fit.}
\label{fig:dimer-fit-errors}
\end{figure}

We first fit to MP2, since both energies and forces are available to achieve a
high-quality fit.  The simplest descriptor used was the distance between pairs
of atoms; each type of pair (e.g. \atompair{C}{C}, \atompair{C}{H}, and
\atompair{H}{H} for methane) is given a separate Gaussian process corresponding
to a separate pair potential.  This descriptor is called `2b' (for ``atomwise
two-body'').  This idea can be extended to triplets of atoms, where the set of
three distances is symmetrized so as to make it permutationally invariant.  This
descriptor is likewise called `3b'.  A first fit was done using both of the
above descriptors, with one Gaussian process for each pair or triplet type; the
resulting potential is essentially a sum of atomwise pair and triplet potentials
with fully flexible functional forms.  This potential is called the `2b+3b GAP'.
As Figure~\ref{fig:dimer-fit-errors} shows, though, this fit offers only a
modest improvement over the baseline.

We therefore attempt a fit in the full six-dimensional space of rigid dimer
configurations using the dimer descriptor.  This descriptor is composed of the
set of distances between all atom pairs in the dimer, symmetrized over
permutations of like atoms.  Concretely, the kernel or covariance function
between two dimers is, as described in~\cite{Bartok2013}:
\begin{equation*}
    k(\bvec{R}, \bvec{R}') = \delta \exp\left[ -\sum_i \frac{(R_i -
    R_i')^2}{(2\sigma_i^2)}\right],
\end{equation*}
where $\bvec{R}$ is the set of distances between all atoms in the dimer,
$\delta$ is the characteristic energy scale of variation of the function, and
the $\sigma_i$ are the charateristic length scales for each distance type.  This
kernel must be permuationally symmetrized so that the resulting potential does
not depend on the order of the atoms:
\begin{equation*}
    \tilde{k}(\bvec{R}, \bvec{R}') = \frac{1}{|S|} \sum_{\pi \in S}
    k(\pi(\bvec{R}), \bvec{R}')
\end{equation*}
where $S$ is the permutation group of the methane dimer, which -- allowing both
swaps of hydrogen atoms within the monomers and swaps of whole monomers in the
dimer -- has order $4! \times 4! \times 2 = 1152$.  The kernel is finally
multiplied by a cutoff function $f_\text{cut}(r_{ab})f_\text{cut}(r'_{ab})$, one
for each dimer, which depends on the center-of-mass separation of the monomers
in the dimer.  The cutoff function is designed to take the function smoothly to
zero as either of the dimers approaches some cutoff distance; in our
implementation, it takes the form of a half-cosine between an inner and an outer
cutoff; the functional form is given in~\cite{Bartok2015a}.

It is an overcomplete representation of the full space of mutual dimer
orientations, which in the case of rigid methanes is six-dimensional.  The
resulting fit offers improvements of at least an order of magnitude across the
entire close range (\SIrange{3}{6}{\angstrom}).  The final correction is the
difference from MP2 to coupled cluster CCSD(T)-F12, which is easily captured to
high accuracy using an atomwise two-body (pairwise) GAP fit to the original
sample of 896 dimers (a subset of the full sample, without the subsequent
short-range augmentation needed for the MP2 dimer GAP).
The composite model created by adding the \ljbaseline, the MP2 dimer GAP, and
the final coupled-cluster two-body GAP, will hereafter be referred to as the
`\gendimgap'.  The resulting potential is a pairwise-additive two-body model and
will thus miss all beyond-two-body (beyond-dimer) effects.  It will still serve
as a useful reference for further models, though, as it can be taken as the
benchmark standard for the fictitious system of methane with only two-body
interactions present.

The new potential was evaluated on its own training set and on the dimer binding
curves from the main text.  It consistently achieves the level of accuracy
specified in the fit, \SI{2}{\milli\electronvolt}, in the regions of the
potential probed under liquid conditions (as evidenced by the pair correlation
function) and can therefore be used as a reference standard for liquid methane
dimer interactions.


\section{GAP fits and evaluation}

GAP is a statistical learning method and hence the quality of the SOAP fits can
be evaluated by how well they reproduce the energies and forces of the training
set.  RMS energy and force errors are given in Table~\ref{tab:gap-errors}, with
the 6-D dimer GAP errors given for comparison.

\begin{table}[ht]
\begin{tabular}{l S[table-format=3.1] S[table-format=1.2]}
    \toprule
    GAP name & {RMS energy error / (\si{\micro\electronvolt\per\methane})} &
        {RMS force error / (\si{\milli\electronvolt\per\angstrom})} \\
    \colrule
    PBE SOAP-GAP          & 200  & 7.36 \\
    PBE0 SOAP-GAP         & 207  & 6.58 \\
    MBD(PBE) SOAP-GAP     & 76.1 & 1.3\text{ (FD)} \\
    MBD(PBE0) SOAP-GAP    & 76.9 & {\textemdash} \\
    MP2 dimer GAP         & 367  & 5.10\\
    CCSD(T)-F12 2b GAP    & 80.6 & {\textemdash} \\
    6-D dimer GAP         & 381  & {\textemdash} \\
    \botrule
\end{tabular}
\caption{RMS energy and force training errors of the GAP fits}
\label{tab:gap-errors}
\end{table}

\begin{figure}[t]
    \centering
    \includegraphics[scale=0.9]{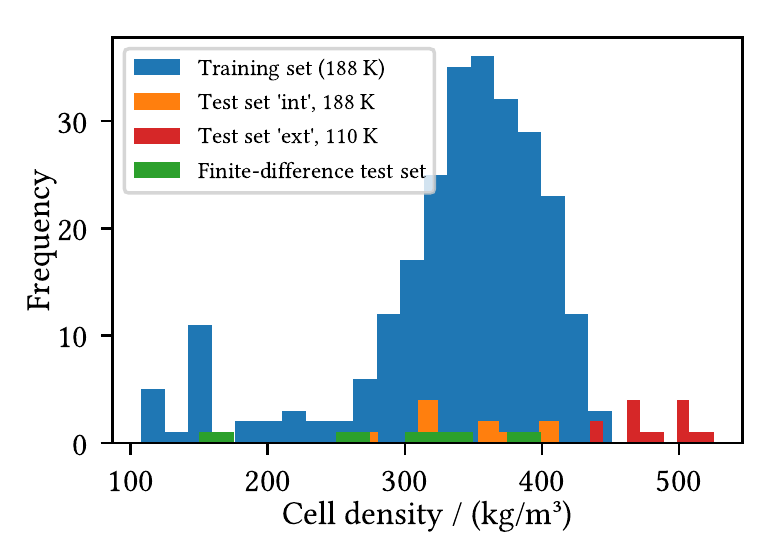}
    \caption{Histogram of densities in the training set and the three test sets:
             Interpolation, extrapolation, and the five finite-difference
             geometries.}
    \label{fig:density-histograms-withfd}
\end{figure}

The fits are additionally evaluated on two test sets that were not included in
the training set, as described in the main text.  Additionally, the performance
of the MBD SOAP-GAP fits were assessed by the finite-difference method, as
gradients were not available.  For this purpose, a sample of five small cells
containing eight methane molecules each was taken from OPLS/AMBER NVT simulations,
one each at five densities ranging from \SIrange{150}{400}{\kgpcm} (see
Figure~\ref{fig:density-histograms-withfd}).  Each geometry
was displaced in each of five randomly selected directions for a total of 25
finite-difference forces.  The results are shown in the right-hand panels of
Figure~\ref{fig:mbdgap-corr}.

\renewcommand{\floatpagefraction}{0.7}

\begin{figure}[p]
    \centering
    \includegraphics[scale=0.9]{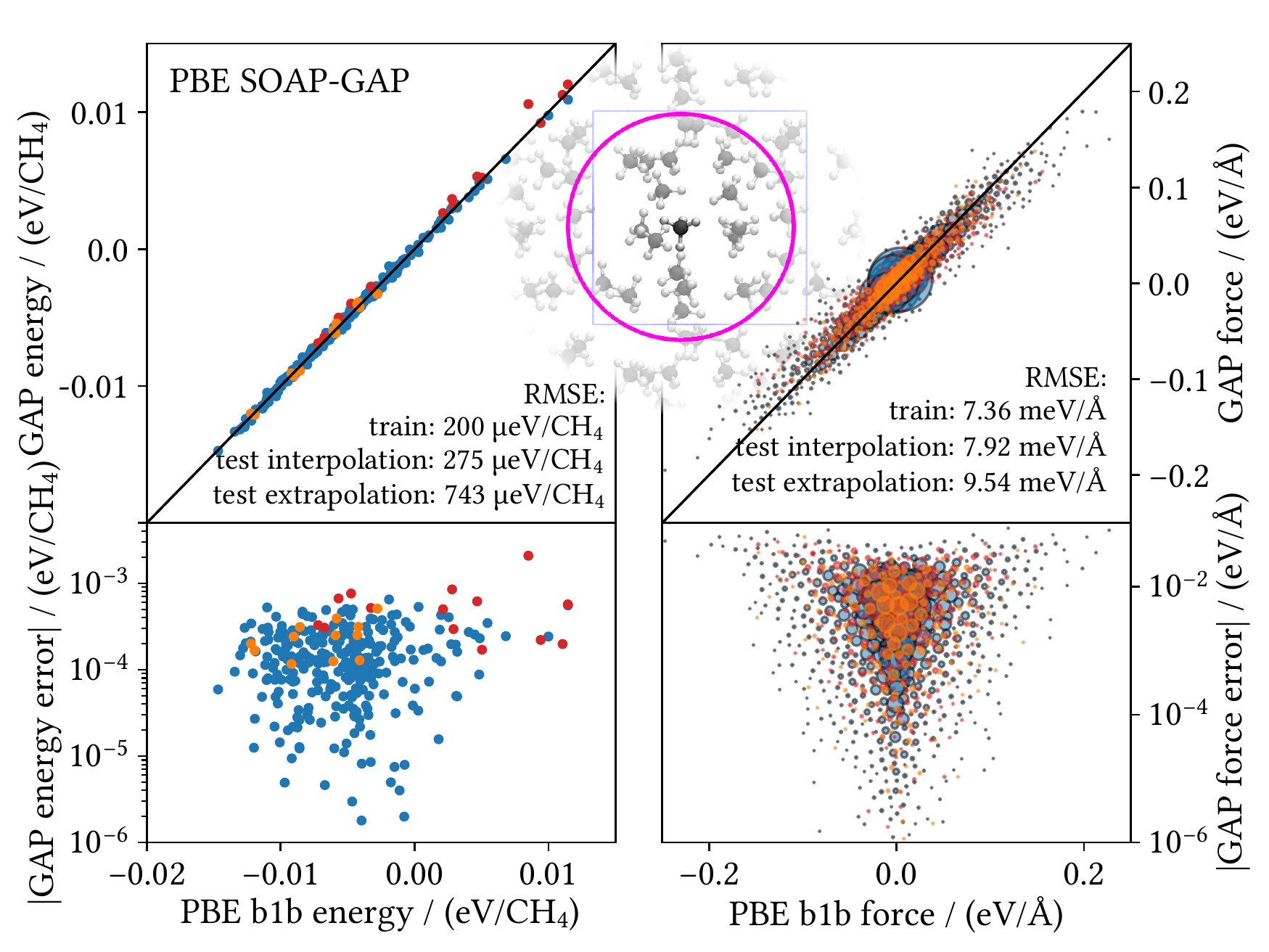}
    \caption{The PBE SOAP-GAP fit on 277 cell
        interaction (beyond one-body, `b1b') energies and corresponding forces.
        Left: energies; right: Cartesian force components.  In the
        force plots, due to the large number of points, only a subset of
        representative points are shown with their sizes scaled according to the
        0.8 power of the number of points they represent.  Top: Correlation
        plots with the line $y=x$ of perfect correlation.  Bottom: Errors on a
        logarithmic scale.  The blue points represent the training set.  The
        orange points represent the interpolation test set and the red points
        represent the extrapolation test set (color online), neither of which
        was used in training the model.}
    \label{fig:pbe0gap-corr}
\end{figure}

\begin{figure}[p]
    \centering
    \includegraphics[scale=0.9]{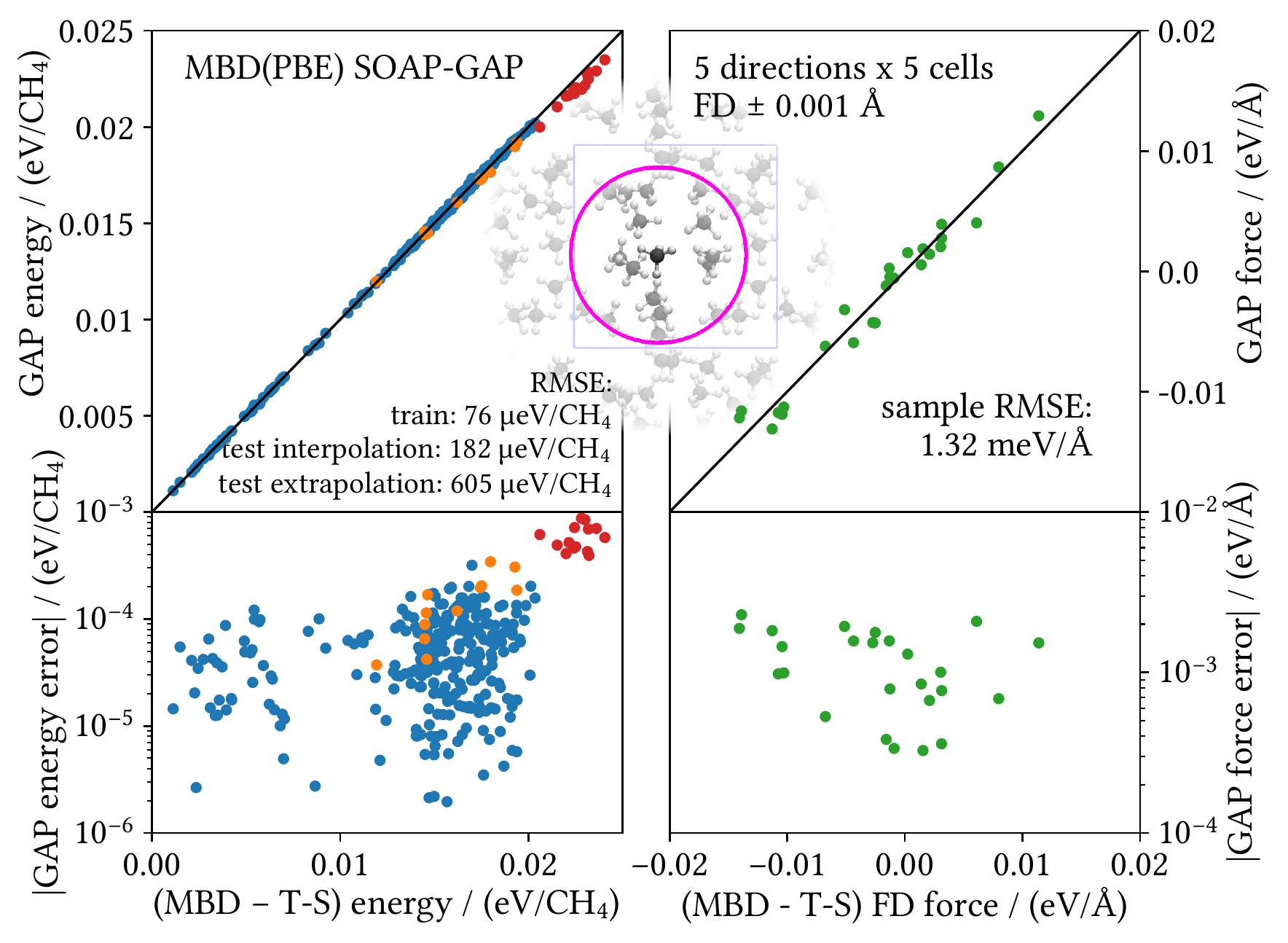}
    \caption{The MBD(PBE) SOAP-GAP fit to the differences between the MBD
        interaction energy and the T-S interaction energy, both computed using
        PBE-derived Hirshfeld volumes (fixed averages for T-S).  Left: Energies;
        blue points for the training set, orange points for the interpolation
        test set and red points for the extrapolation test set (color online).
        Right: Forces estimated by finite differences on five cells of eight
        methanes each in five randomly chosen directions each (green points).
        As before, correlation plots against $y=x$ above, log errors below.}
    \label{fig:mbdgap-corr}
\end{figure}

The parameters for the above fits are given as command lines that can be used
with the `\verb+teach_sparse+' command in the libAtoms/QUIP
package~\cite{libAtoms}\todo{needs updating}.  The GAP code can be downloaded at
\url{http://www.libatoms.org/gap/gap_download.html}, with a prepackaged version
available through Docker at \url{https://hub.docker.com/r/libatomsquip/quip/}.


The parameters for the PBE SOAP-GAP are (all on one line):
\begin{Verbatim}[samepage=true]
teach_sparse at_file=mebox-minimal-nots-b1b-train.xyz gap={
      soap atom_sigma=0.5 l_max=8 n_max=8 cutoff=6.0
      cutoff_transition_width=1.0 delta=0.01
      add_species n_species=2 species_z={{1 6}}
      n_sparse=2000 covariance_type=dot_product sparse_method=cur_points
      zeta=4.0
    } default_sigma={0.0001 0.002 1.0 1.0} sparse_jitter=1e-10
    virial_parameter_name=none gp_file=gp-mebox-pbe-b1b.xml 
\end{Verbatim}
(The parameters for the PBE0 SOAP-GAP are exactly the same; only the source data
was computed with PBE0 instead of PBE)

The parameters for the MBD SOAP-GAP are (again, same for both PBE and PBE0
energies):
\begin{Verbatim}[samepage=true]
teach_sparse at_file=mebox-minimal-mbdint.xyz
    core_param_file=../python/dispts_quip_params.xml core_ip_args={
        Potential xml_label=ts
        calc_args={hirshfeld_vol_name=hirshfeld_avg_volume}
    } e0=0.0 gap={
        soap atom_sigma=0.5 l_max=8 n_max=8 cutoff=5.0
        cutoff_transition_width=1.0 delta=0.001
        add_species n_species=2 species_z={{1 6}} n_sparse=2000 
        covariance_type=dot_product sparse_method=cur_points zeta=4.0
    } default_sigma={0.0001 1.0 1.0 1.0}
    sparse_jitter=1e-10 gp_file=gp-mbd-soap.xml
\end{Verbatim}

\pagebreak[4]

The parameters for the 6-D dimer fit to MP2 are:
\begin{Verbatim}[samepage=true]
teach_sparse at_file=me-rigid-shortaug3-mp2-avqz-intnonan.xyz
    core_param_file={../empirical-pots/ljrep_quip_params.xml}
    core_ip_args={IP LJ} gap={
        general_dimer cutoff=6.0 cutoff_transition_width=1.0
        signature_one={{6 1 1 1 1}} signature_two={{6 1 1 1 1}}
        monomer_one_cutoff=1.5 monomer_two_cutoff=1.5 atom_ordercheck=F
        strict=F mpifind=T theta_uniform=1.0 covariance_type=ARD_SE
        n_sparse=2000 delta=0.02 sparse_method=CUR_COVARIANCE
    } default_sigma={0.0002 0.002 0.0 0.0} sparse_jitter=1e-10
    energy_parameter_name=energy force_parameter_name=force e0=0.0
    gp_file=gp-merig-mp2-gendim-shortaug3.xml do_copy_at_file=F
\end{Verbatim}

And for the much simpler two-body (atomwise) fit to the CCSD(T)-MP2 difference:
\begin{Verbatim}[samepage=true]
teach_sparse at_file=me-rigid-train-ljrep.xyz gap={
      distance_2b cutoff=10.0 covariance_type=ARD_SE
      n_sparse=50 sparse_method=UNIFORM Z1=6 Z2=6
      theta_fac=0.2 delta=0.0005 resid_name=resid only_inter=T
      : distance_2b cutoff=6.0 covariance_type=ARD_SE
      n_sparseX=50 sparse_method=uniform Z1=1 Z2=6
      theta_fac=0.2 delta=0.0005 resid_name=resid only_inter=T
      : distance_2b cutoff=6.0 covariance_type=ARD_SE
      n_sparseX=50 sparse_method=uniform Z1=1 Z2=1
      theta_fac=0.2 delta=0.0005 resid_name=resid only_inter=T
    } default_sigma={0.00001 0.0 0.0} sparse_jitter=1e-10
    energy_parameter_name=ediff_cc force_parameter_name=none e0=0.0
    gp_file=gp-merig-cc-ljrep-2b.xml do_copy_at_file=F sparse_separate_file=F 
\end{Verbatim}
The final 6-D dimer GAP is simply the sum of the above two potentials.

These potentials, once fitted, are stored in the form of an XML file that can be
read by QUIP to evaluate energies and forces on any new configuration.  The XML
files for the above GAPs are avalable online in the Apollo
repository~\footnote{\protect\url{https://doi.org/10.17863/CAM.26364}} as well
as on our group's
webpage~\footnote{\protect\url{http://www.libatoms.org/Home/DataRepository}}.

\section{DFT and MBD parameters}

As mentioned in the main text, the sample for the DFT calculations was taken
from MD trajectories under liquid conditions run using a classical potential
(OPLS/AMBER~\cite{Jorgensen1988,amber}) at a temperature of \SI{188}{\kelvin}
and five pressures ranging from \SIrange{0}{400}{\bar} (the same ones at which
OPLS/AMBER was tested in the main text, with the addition of \SI{0}{\bar}).
There were 60 samples taken from each pressure, with the exception of the
shorter \SI{0}{\bar} simulation, which only contributed 40 samples.  Each of the
280 cells in the sample contained 27 (flexible) methane molecules; otherwise,
the simulation parameters were the same as those described later in
Section~\ref{sub:md-analytical}.

The DFT calculations were all done with the CASTEP code~\cite{CASTEP}, version
8.0.  The PBE calculations were done with a plane-wave cutoff of
\SI{650}{\electronvolt} and the default finite-basis correction.  Due to the
large, amorphous nature of the system, no k-point sampling was employed;
calculations were only done at the $\Gamma$ point.  Convergence tolerances were
set to \SI{1}{\micro\electronvolt\per\atom} for the energies and
\SI{10}{\micro\electronvolt\per\angstrom} for the forces.  The PBE0 calculations
were done with a cutoff of \SI{700}{\electronvolt} and no finite-basis
correction, again only at the $\Gamma$ point, and convergence tolerances one
order of magnitude smaller (\SI{0.1}{\micro\electronvolt\per\atom} for the
energies and \SI{1}{\micro\electronvolt\per\angstrom} for the forces).  Since
computing the interaction energy requires subtracting the one-body contribution
(the energy and force of each individual methane molecule in the cell) and the
samples had flexible monomer geometries, an additional calculation was run on
each of the 27 individual molecules in each cell, using the same periodic
boundary conditions as the full cell.  The energy that resulted from subtracting
the sum of the monomer energies from the total cell energy is the interaction or
beyond-one-body (`b1b') energy (and likewise with the interaction force).
Finally, two cells were discarded because their interaction energies (both PBE
and PBE0) were much higher than the rest; those cells came from the initial MD
equilibration from a high-energy geometry, so they were removed to achieve a
better fit for normal, equilibrium conditions.  Additionally, the largest cell
for PBE and the largest 20 cells for PBE0 did not complete because the
computational requirements exceeded available resources.
The training set therefore comprised 277 PBE interaction energies (and $277
\times 135 \times 3 = 112185$ PBE force components), and 258 PBE0 interaction
energies (and $258 \times 135 \times 3 = 104490$ PBE0 force components).  These
sets of interaction energies and forces were finally fit with the SOAP GAPs
above, ranged at \SI{6}{\angstrom}.  

The MBD energies were computed on the same sets of 277 (or 258 for PBE0) methane
cells using the
implementation available at \url{http://www.fhi-berlin.mpg.de/~tkatchen/MBD/}
and interfaced with QUIP\todo{make the patched MBD version available}.  This was
done both with the PBE and PBE0 Hirshfeld volumes calculated from each geometry,
as reported by CASTEP.  The supercell cutoff parameter was adjusted so that a
$1 \times 1 \times 1$ supercell (that is, only the unit cell) was used, in
correspondence with the omission of k-point sampling in the DFT calculations.
All other MBD paramters were left at their defaults.  The corresponding T-S
model, with fixed, per-element averaged PBE or PBE0 volumes, was then subtracted
and the difference was fit with a SOAP-GAP ranged at \SI{5}{\angstrom}.  The
magnitude of the correction beyond this range was small enough that neglecting
it was seen as safe.  The given implementation did not implement gradients, so
the accuracy of the GAP forces was assessed using a finite-difference scheme as
described above.

The Hirshfeld volumes used to compute T-S and MBD energies on the dimer test set
(computed to assess PBE+MBD and PBE0+MBD dimer model errors) were instead
computed from the wavefunctions produced by the \textsc{Psi4} code~\cite{psi4}
with the HORTON post-processing functionality~\cite{horton}, which itself uses
methods derived by Becke and Dickson for polyatomic
molecules~\cite{becke1988_multicenter,becke1988_poisson,lebedev1999}.

\section{MD parameters}

\subsection{GAP fits}
\begin{figure}
    \includegraphics[scale=0.9]{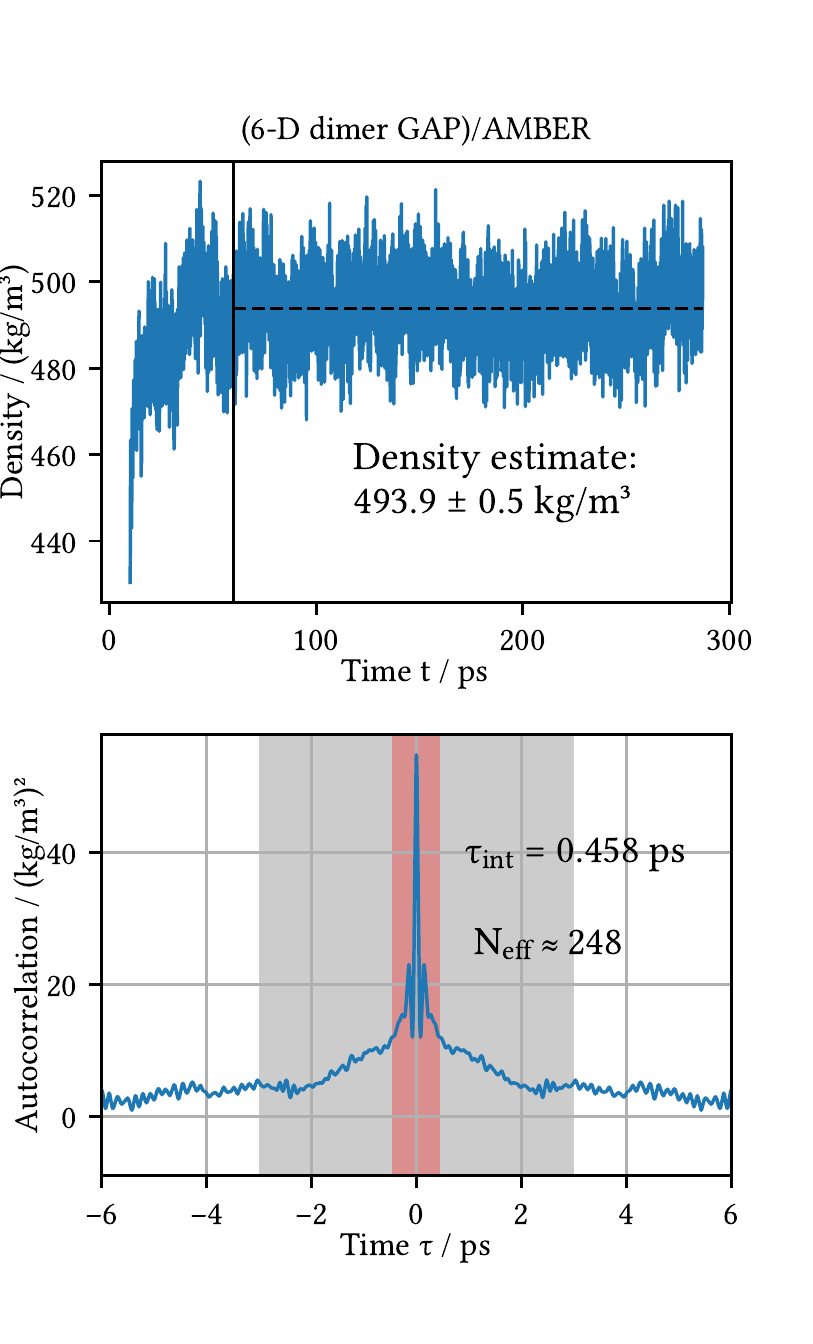}\quad
    \includegraphics[scale=0.9]{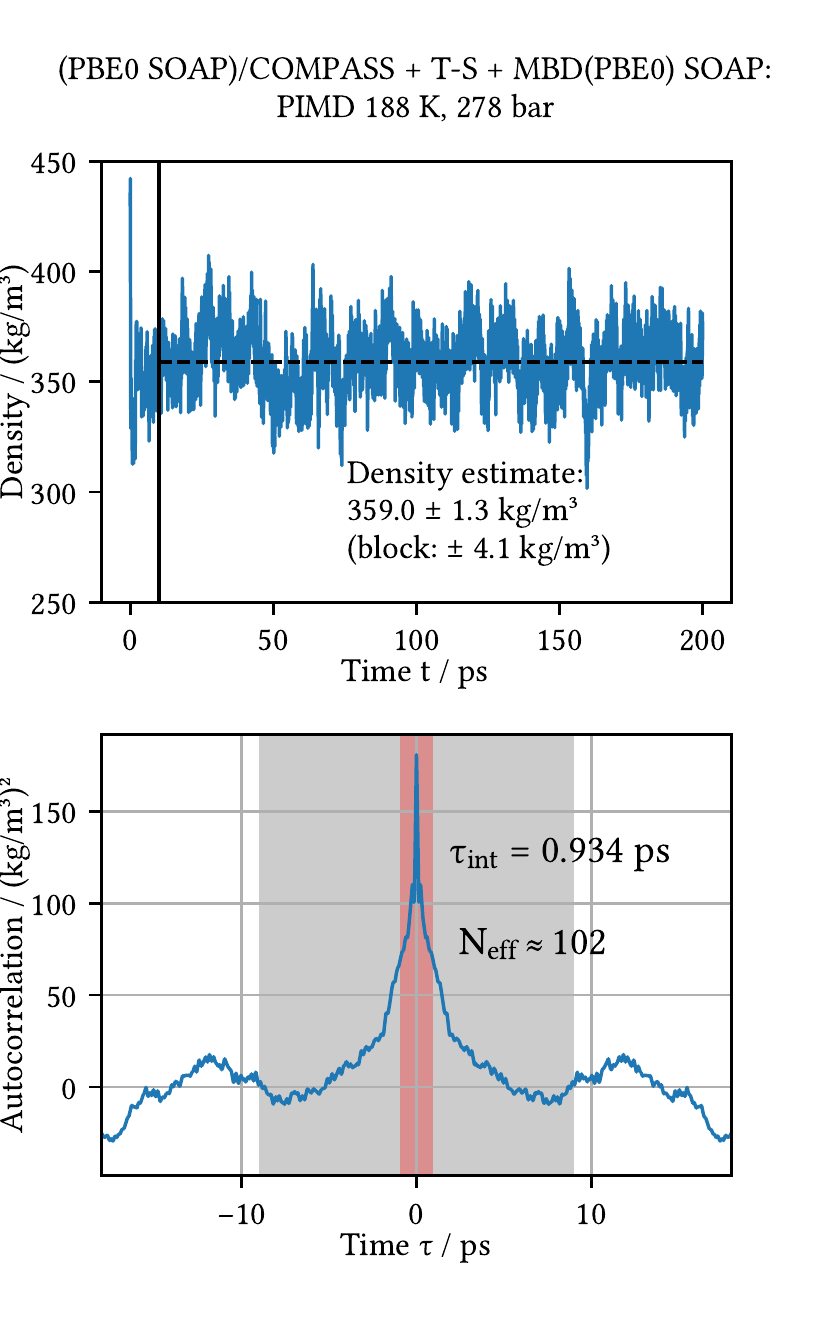}
    \caption{Trace of the density of the GAP NPT simulations over time.
        Averaging was done starting from the vertical solid line.
        The autocorrelation of the density timeseries (discarding the initial
        transient) is shown in the bottom plots.  The integral
        $\tau_\text{int}$ of the normalized autocorrelation is a good estimate
        for the series's correlation time, which in turn can be used to estimate
        the number of \emph{effective} independent samples $N_\text{eff} =
        \frac{T}{2\tau_\text{int}}$ ($T$ is the length of the series being
        averaged) and the standard error on the mean $\sigma_\text{corr} =
        \sqrt{\frac{\sigma_0^2}{2 N_\text{eff}}}$ (where $\sigma_0^2$ is the
        variance of the sample being averaged).  The grey region is the
        integration region and the red shows the correlation time. }
    \label{fig:timeseries0}
\end{figure}

\begin{figure}
    \includegraphics[scale=0.9]{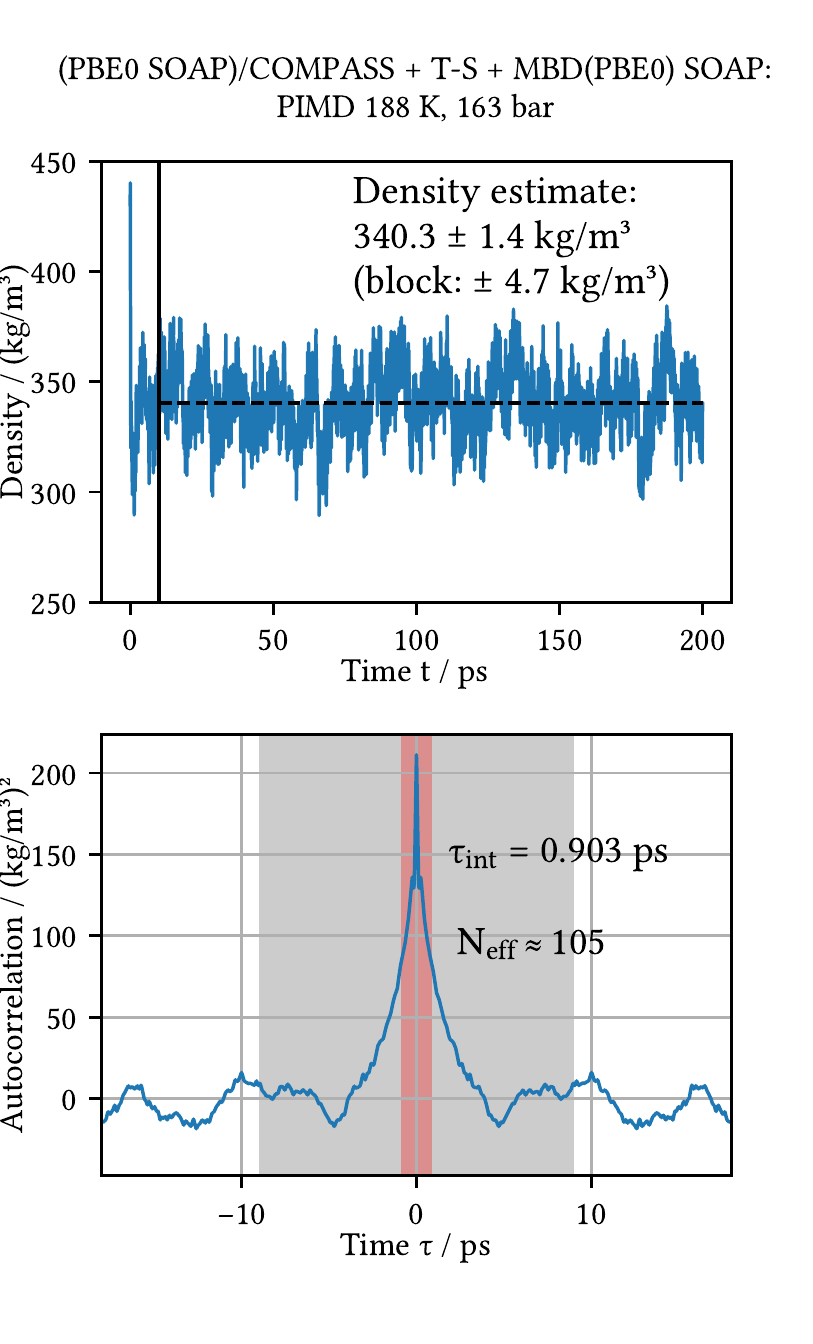}\quad
    \includegraphics[scale=0.9]{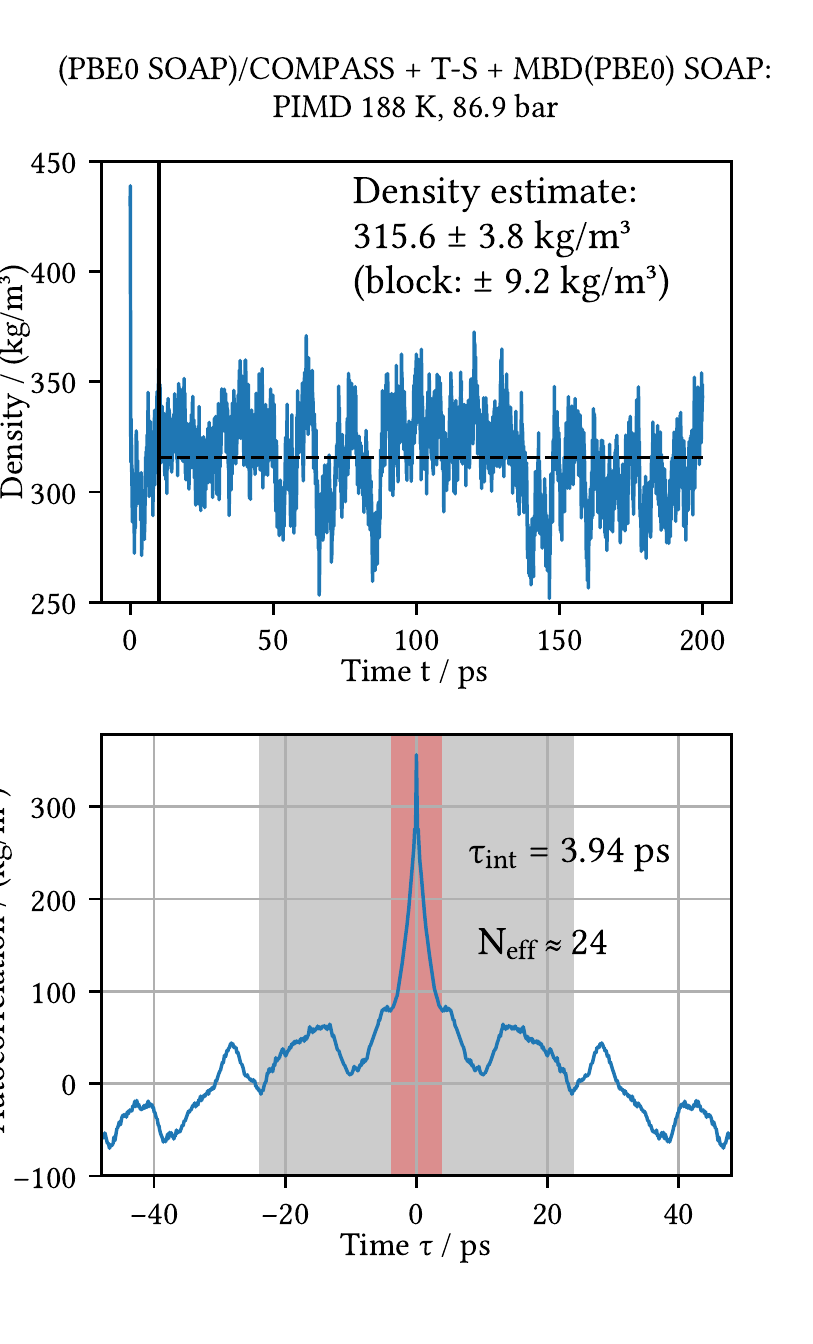}
    \caption{Trace of the density of the GAP NPT simulations over time:
        (PBE0 SOAP)/COMPASS + T-S + MBD(PBE0) SOAP at \SI{188}{\kelvin}.
        As in Figure~\ref{fig:timeseries0}, timeseries at the top,
        autocorrelation plots at the bottom.  Most of the PIMD
        simulations are not yet completely equilibrated (as can be seen both in
        the time trace and the autocorrelation function), so an interim
        estimate was also computed by splitting the utilized simulation time
        into 10 blocks and taking the standard deviation of the individual means
        of the blocks. }
\end{figure}

\begin{figure}
    \includegraphics[scale=0.9]{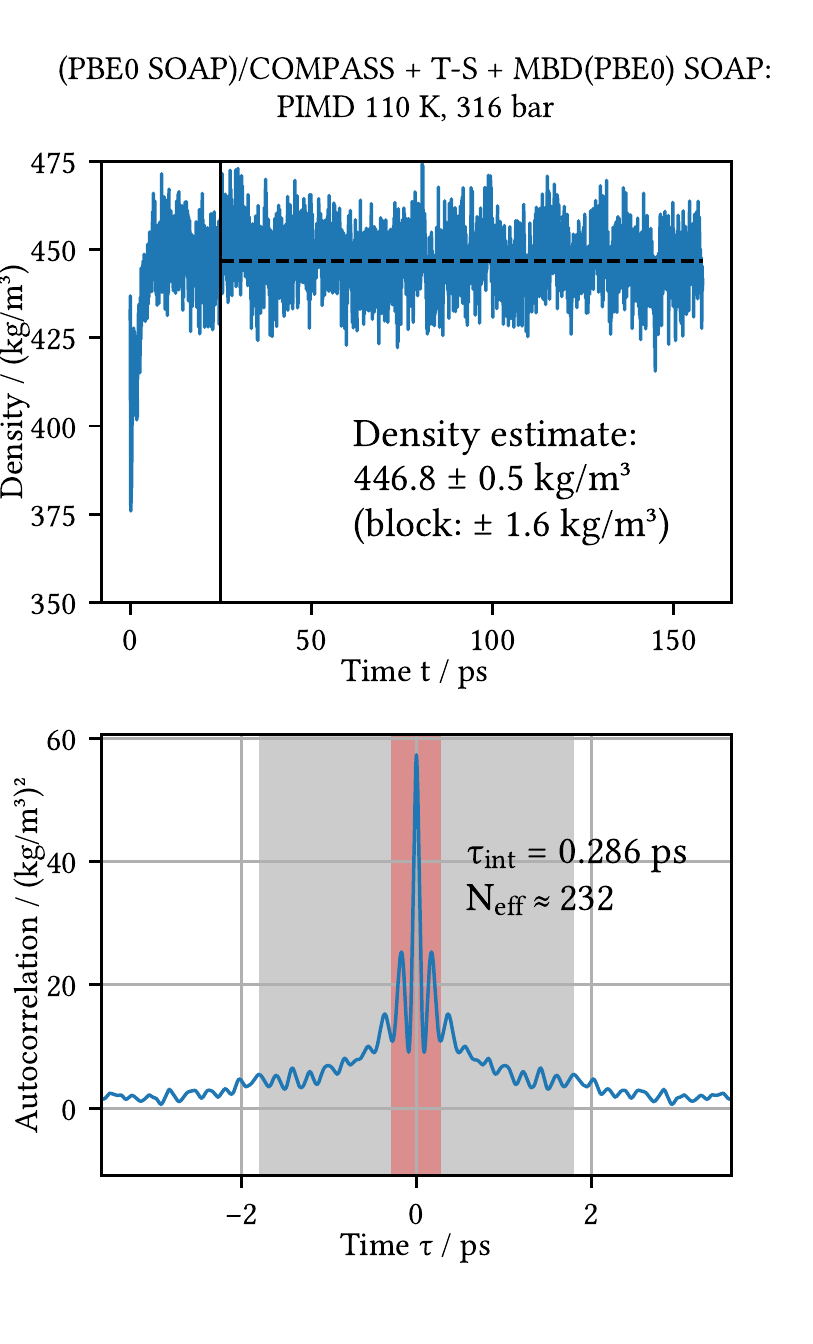}\quad
    \includegraphics[scale=0.9]{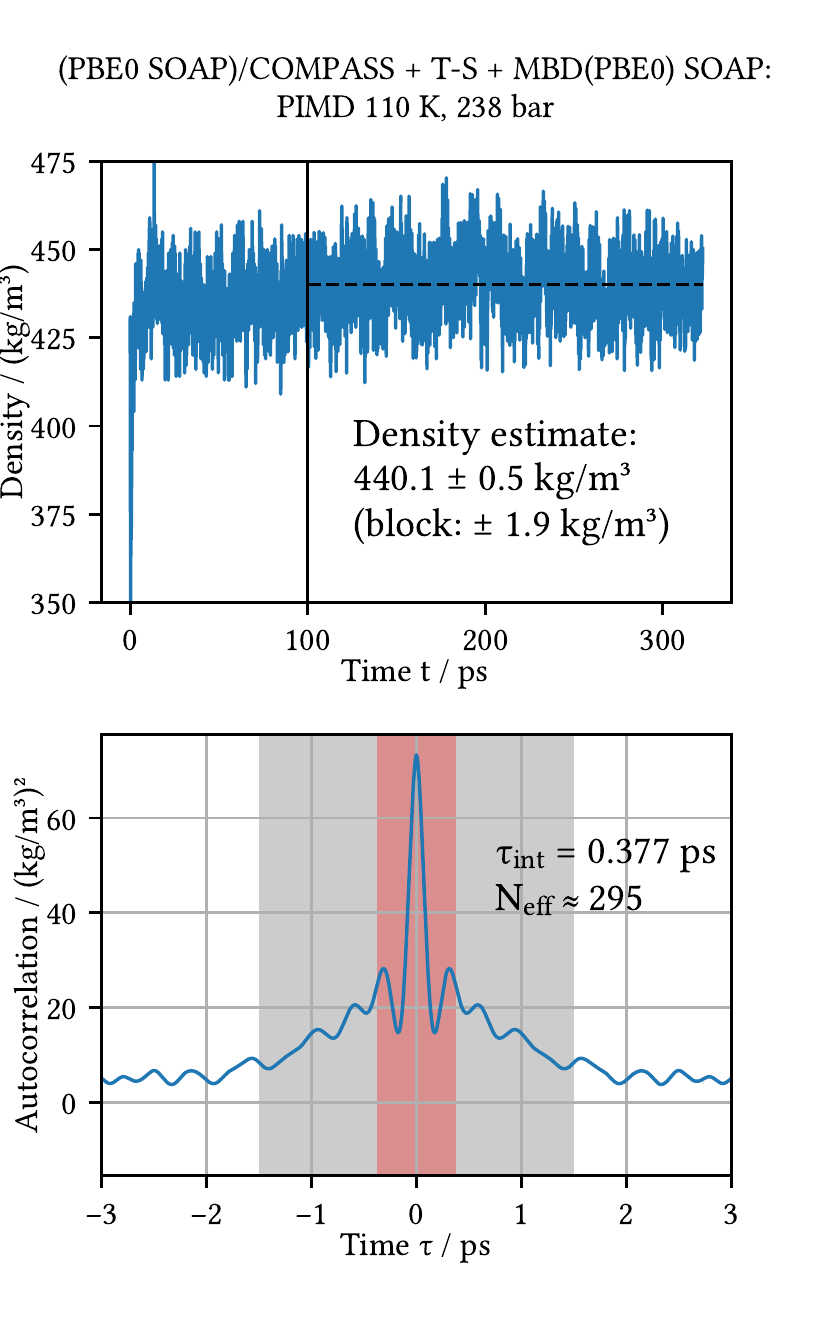}
    \caption{Trace of the density of the GAP NPT simulations over time:
        (PBE0 SOAP)/COMPASS + T-S + MBD(PBE0) SOAP at \SI{110}{\kelvin}.
        As in Figure~\ref{fig:timeseries0}, timeseries at the top,
        autocorrelation plots at the bottom.}
\end{figure}

\begin{figure}
    \includegraphics[scale=0.9]{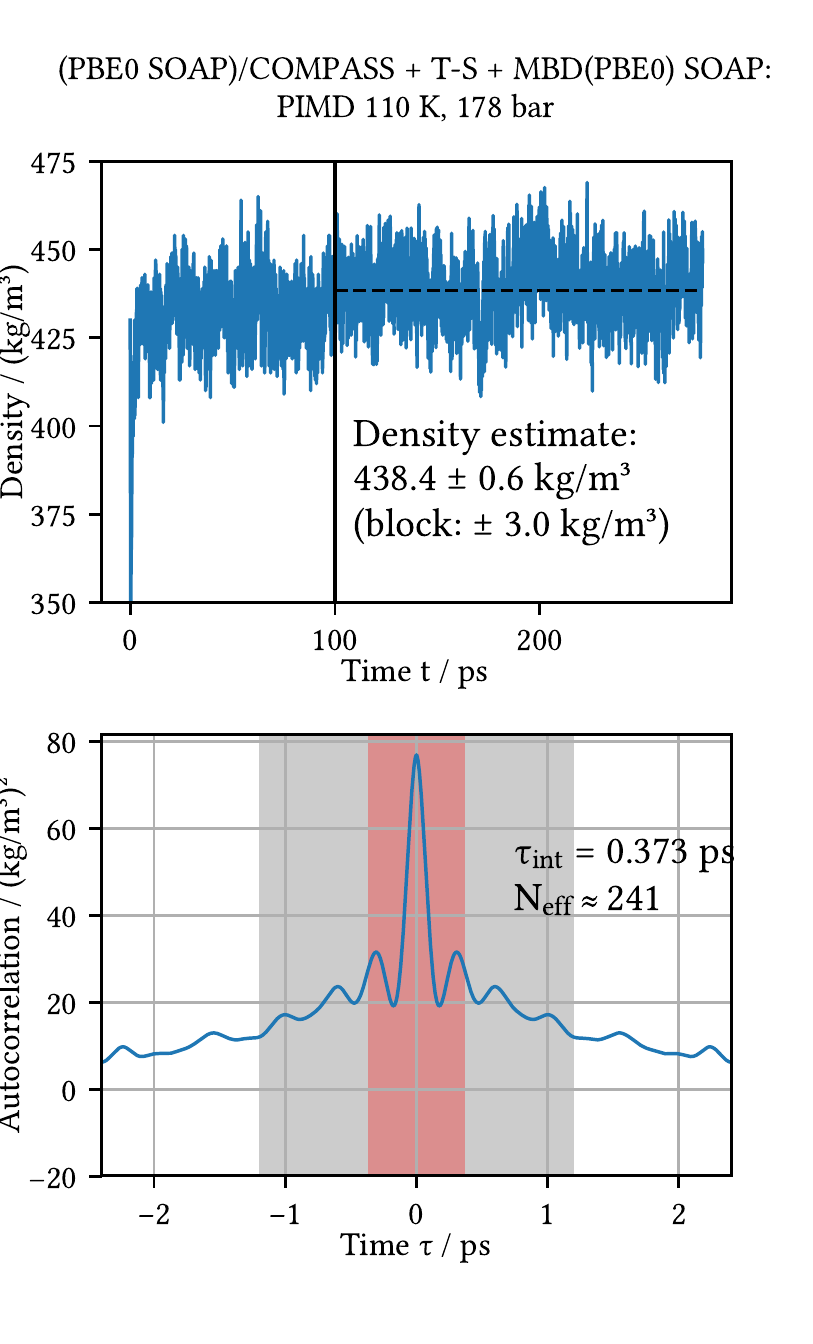}\quad
    \includegraphics[scale=0.9]{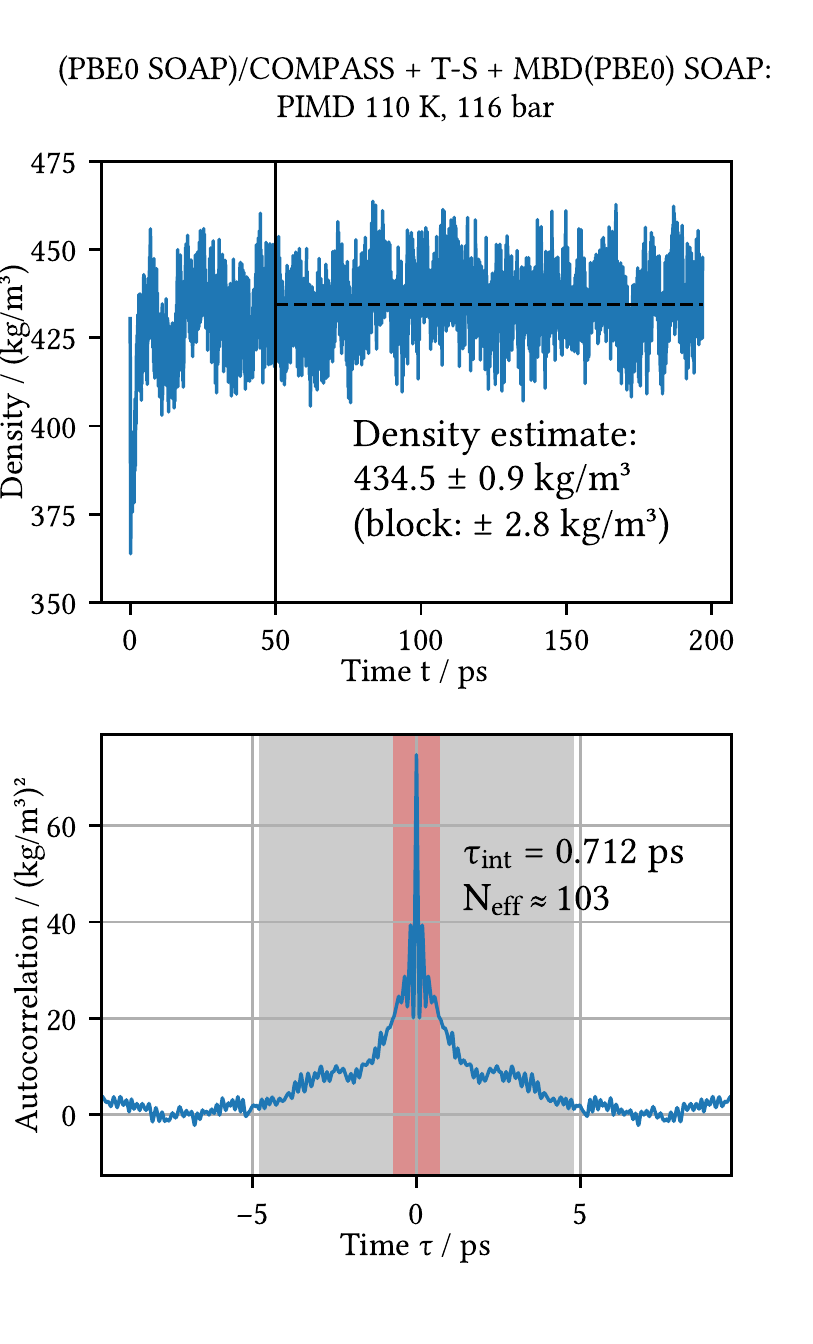}
    \caption{Trace of the density of the GAP NPT simulations over time:
        (PBE0 SOAP)/COMPASS + T-S + MBD(PBE0) SOAP at \SI{110}{\kelvin}.
        As in Figure~\ref{fig:timeseries0}, timeseries at the top,
        autocorrelation plots at the bottom.}
\end{figure}

\begin{figure}
    \includegraphics[scale=0.9]{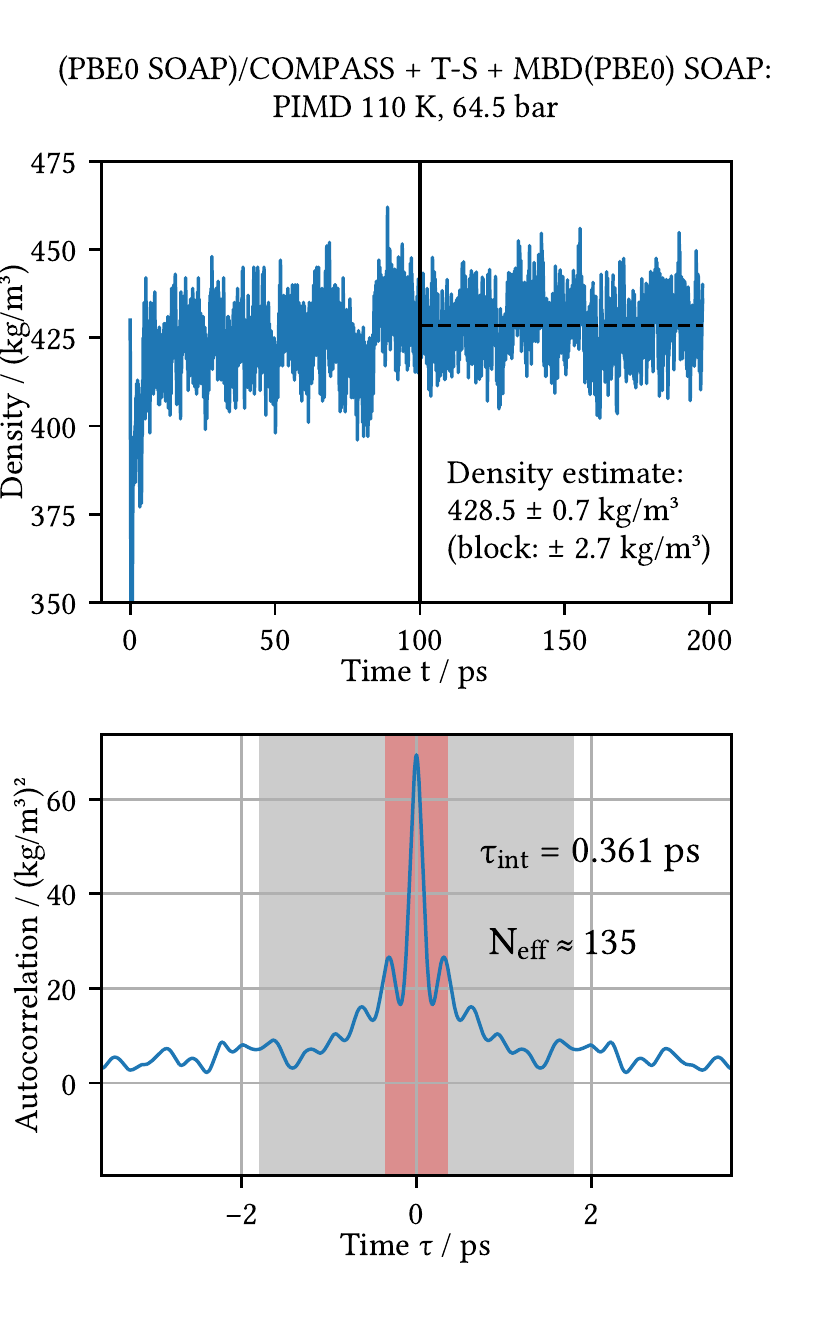}\quad
    \includegraphics[scale=0.9]{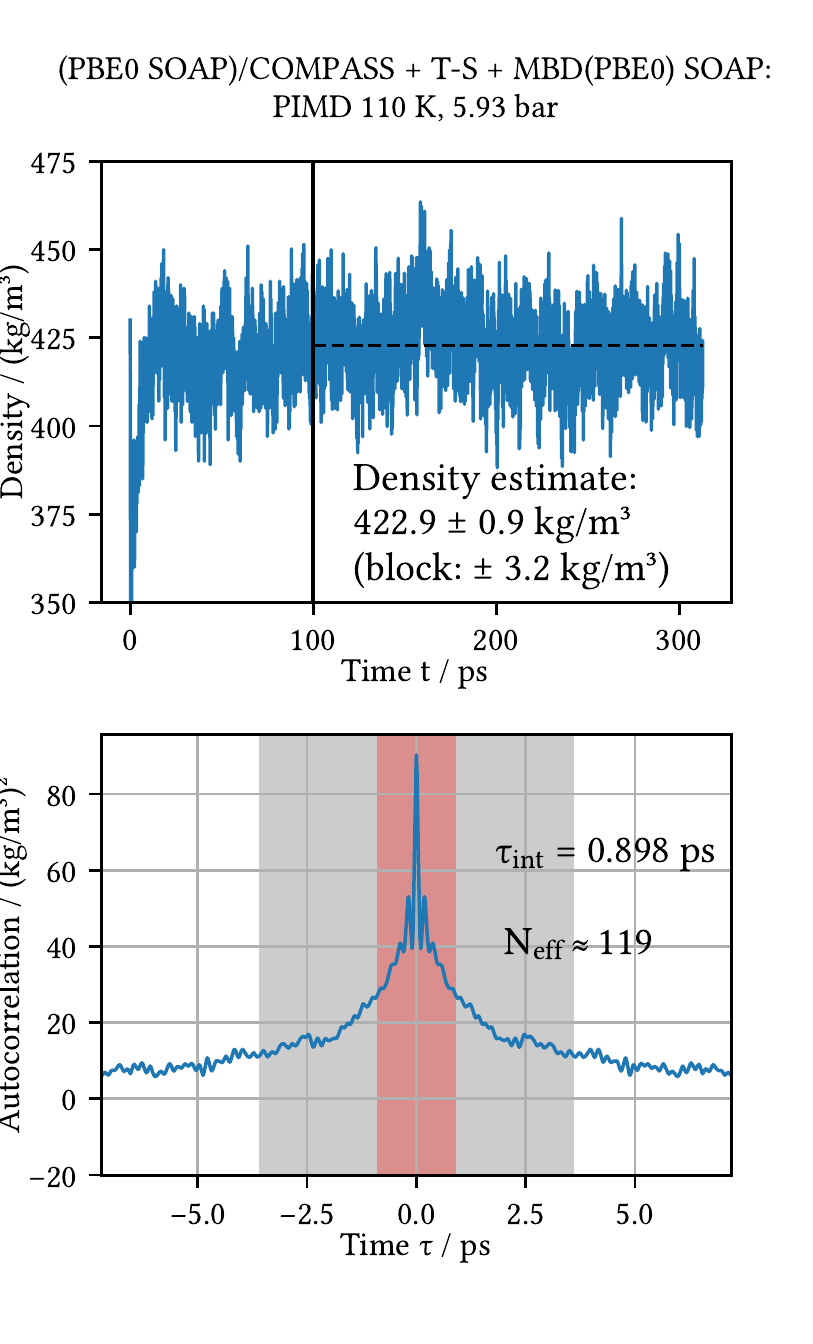}
    \caption{Trace of the density of the GAP NPT simulations over time:
        (PBE0 SOAP)/COMPASS + T-S + MBD(PBE0) SOAP at \SI{110}{\kelvin}.
        As in Figure~\ref{fig:timeseries0}, timeseries at the top,
        autocorrelation plots at the bottom.}
    \label{fig:timeseries-1}
\end{figure}

Most of the GAP MD simulations were run with i-PI~\cite{Ceriotti2014},
interfaced through LAMMPS~\cite{lammps,lammps11Aug17} to QUIP~\cite{libAtoms}.
Only the 6-D dimer GAP simulation was run with QUIP's built-in MD functionality.
It used the adaptive Langevin thermostat of Jones and
Leimkuhler~\cite{Jones2011} (with a time constant of \SI{10}{\femto\second}) and
a Hoover-Langevin barostat~\cite{Quigley2004} (with a time constant of
\SI{100}{\femto\second} and a mass factor of 100).  For the SOAP simulations,
the T-S correction was cut off at \SI{15}{\angstrom}, smoothed with a
half-cosine curve over \SI{1}{\angstrom}.  Likewise, the \ljbaseline\ (as a
component of the general dimer GAP) was cut off at \SI{15}{\angstrom} and
smoothed over \SI{1}{\angstrom} (\atompair{C}{C} potential only; the other two
were simply cut off at \SI{10}{\angstrom}).  Analytical tail corrections were
calculated by computing the integral of the missing energy and virial outside
the inner cutoff of \SI{14}{\angstrom}; see Section \ref{sec:tail-integral} for
details.  The initial configuration for these simulations was a 100-methane cell
generated using Packmol~\cite{Martinez2009}.

The i-PI simulations used a thermostat based on
the generalized Langevin equation (GLE, otherwise known as colored-noise
thermostats), namely the ``smart sampling'' method of Ceriotti, Bussi, and
Parrinello~\cite{Ceriotti2010}.  The parameters were generated at
\mbox{\url{http://gle4md.org/}} using the parameters $t_\text{opt} =
{}$\SI{10}{ps}, $N_s = 6$, and $\sfrac{\omega_\text{max}}{\omega_\text{min}} =
10^4$ for the thermostat and $t_\text{opt} = {}$\SI{2}{\pico\second}, $N_s = 6$,
and $\sfrac{\omega_\text{max}}{\omega_\text{min}} = 10^3$, and a piston time
constant of $\tau = {}$\SI{100}{\femto\second} for the barostat.

The PIMD simulations used the PIGLET~\cite{Ceriotti2009a,Ceriotti2012}
thermostat to accelerate convergence to the quantum partition function.  The
thermostat parameters were generated at the same website, this time using the
PIGLET paramters of OPT(H), $N_s = 8$, $\omega_\text{max} =
{}$\SI[per-mode=reciprocal]{3000}{\per\cm},
$\sfrac{\omega_\text{max}}{\omega_\text{min}} = 10^4$, $\sfrac{\hbar\omega}{k_B
T} = 50$, with the appropriate temperature $T$ (\SI{110}{\kelvin} or
\SI{188}{\kelvin}), and with both 12 and 16 beads to verify convergence to the
quantum limit; the larger number was used in production simulations.  The
centroid barostat used the analogous ``optimal sampling''
method~\cite{Ceriotti2009,Ceriotti2010} with the same parameters: Potential
energy optimized, $N_s = 8$, $\omega_0 =
{}$\SI[per-mode=reciprocal]{30}{\per\cm}, and
$\sfrac{\omega_\text{max}}{\omega_\text{min}} = 10^4$ (resulting in
$\omega_\text{max} = {}$\SI[per-mode=reciprocal]{3000}{\per\cm}).

The parameter files for the above thermostats, including the matrices used to
propagate the generalized Langevin equation, are available in the Apollo
repository~\cite{Note1} as well as on our group's webpage~\cite{Note2}.

No smooth cutoffs were done in the i-PI simulations due to the necessity of
interfacing with QUIP through LAMMPS; analytical tail corrections were applied,
though.  The initial configuration was prepared with an initial
\SI{10}{\pico\second} NVT equilibration using the `(PBE0 SOAP)/COMPASS + T-S +
MBDGAP' potential; this configuration was used for both the classical and PIMD
simulations at that temperature.  All GAP MD simulations were done with a
timestep of \SI{0.5}{\femto\second}.

Each run had a certain amount of initial equilibration time discarded from its
trajectory, depending chiefly on the potential, the thermostat, and the
temperature.  The ten pairs of plots in Figures~\ref{fig:timeseries0}
through~\ref{fig:timeseries-1} show how the
average density and standard error were obtained from the time evolution of the
density for the 6-D dimer GAP simulation and for the `\mbox{(PBE0 SOAP)/COMPASS}
\mbox{+ T-S} \mbox{+ MBD(PBE0) SOAP}' PIMD simulations.  The standard error was
obtained by integrating the autocorrelation of the density timeseries as
described in Sokal~\cite{Sokal1996}.  However, many simulations showed extremely
long correlation times and were not fully equilibrated within the available
simulation time, rendering the autocorrelation method inapplicable.  Therefore,
to estimate the error incurred due to the large-scale fluctuations still
observed, the simulation time utilized for averaging was split into ten equal
blocks (corresponding approximately to the timescale of fluctuations still
observed), the mean value was computed within each of those blocks, and the
final error estimate computed as the standard deviation of those means.  These
error estimates are displayed as error bars on the `\mbox{(PBE0 SOAP)/COMPASS}
\mbox{+ T-S} \mbox{+ MBD(PBE0) SOAP}' PIMD simulation results in Figure~5 of the
main text.

\subsection{\label{sub:md-analytical}Analytical potentials}

The analytical potentials were run in LAMMPS~\cite{lammps,lammps5Oct15} with a Langevin
thermostat~\cite{Bruenger1984} and a Nosé-Hoover
barostat~\cite{Nose1984,Hoover1985,Parrinello1981,Shinoda2004,Tuckerman2006}
with the MTK correction~\cite{Martyna1994}, both using a time constant of
\SI{0.1}{\pico\second}, and an initial configuration of 200 methane molecules
generated using Packmol~\cite{Martinez2009} and relaxed with the
OPLS-AA~\cite{Jorgensen1996} forcefield.  All simulations used analytical tail
corrections to account for the otherwise-neglected dispersion energy beyond
their cutoffs~\cite{Allen+1989,Sun1998} (for the \ljbaseline\ this was only done
for the \atompair{C}{C} potential).  For potentials with a Coulomb component
(OPLS/AMBER and COMPASS), the contributions beyond the cutoff were calculated
with the particle-particle particle-mesh (PPPM) method~\cite{Hockney1988}.
The MD timesteps were \SI{1}{\femto\second} for TraPPE, \SI{0.5}{\femto\second}
for the Li-Chao \lj and OPLS/AMBER at \SI{110}{\kelvin}, and
\SI{0.1}{\femto\second} for the others (OPLS/AMBER at \SI{188}{\kelvin}, the
\ljbaseline, and COMPASS).

The potentials themselves used \lj\ cutoffs (and Coulomb cutoffs for OPLS/AMBER
and COMPASS) of \SI{10}{\angstrom}, except for TraPPE, where the cutoff of
\SI{14}{\angstrom} recommended on the website was used instead.  The two
pairwise \lj\ fits (the \ljbaseline\ and Li-Chao) both were added to the AMBER
intramolecular terms to give complete liquid methane potentials.

Equilibration and run times again varied between the potentials based on the
rate of convergence of the density, although the same times were used throughout
an isotherm.  The times are summarized in Table~\ref{tab:eqtimes-classical}.

\begin{table}[ht]
\begin{tabular}{l S S S}
\toprule
Potential & {Temperature/\si{\kelvin}}
          & {Equilib. time/\si{\pico\second}}
          & {Run time/\si{\pico\second}} \\
\colrule
\multirow{2}{5cm}{\raggedleft TraPPE}
    & 110 & 100 & 400 \\
    & 188 & 100 & 400 \\
\multirow{2}{5cm}{\raggedleft OPLS/AMBER}
    & 110 & 25  & 75 \\
    & 188 & 100 & 400 \\
\multirow{2}{5cm}{\raggedleft \ljbaseline}
    & 110 & 50  & 50 \\
    & 188 & 300 & 200 \\
\multirow{2}{5cm}{\raggedleft Li-Chao L-J}
    & 110 & 100 & 400 \\
    & 188 & 400 & 100 \\
\botrule
\end{tabular}
\caption{Equilibration and run times for the analytical potentials}
\label{tab:eqtimes-classical}
\end{table}

\section{Tail corrections with a smooth cutoff}
\label{sec:tail-integral}

For all the NPT simulations done for this work it was found important to
incorporate tail corrections to account for the missing pressure neglected by
cutting off the long-range dispersion potentials (the sixth-power part of \lj\
and T-S).  These corrections can also be applied in the case where the potential
is smoothed to zero before the cutoff, though the resulting integrals become
more difficult to evaluate.

The expression for the missing pressure in a potential that is cut off with a
smoothing function that starts at $r_\text{in}$ and ends at $r_\text{out}$
is, by straightforward extension of the formulae in~\cite{Allen+1989,Sun1998}:

\begin{align}
    p_\text{exact} - p_\text{cut} &= \frac{1}{6} \sum_{i=1}^{n_\text{typ}} \rho_i
                                \sum_{j=1}^{n_\text{typ}} \rho_j
                    \int_{r_\text{in}}^\infty 
        r \frac{\dee}{\dee r} (\phi_{ij}(r) - \phi_{ij,\text{cut}}(r)) 4 \pi r^2 g_{ij}(r) \dee r \nonumber\\
                 & 
                    \begin{aligned}
               {} = \frac{1}{6} \sum_{i=1}^{n_\text{typ}} \rho_i
                                \sum_{j=1}^{n_\text{typ}} \rho_j
                    & \Bigg[ \int_{r_\text{in}}^{r_\text{out}}
                        r \frac{\dee}{\dee r} \left(\phi_{ij}(r) (1 - S(r))\right)  4 \pi r^2 g_{ij}(r) \dee r \\
                    & {} + \int_{r_\text{out}}^\infty r \frac{\dee \phi_{ij}(r)}{\dee r} 4 \pi r^2 g_{ij}(r) \dee r \Bigg] 
                    \end{aligned}
                    \label{eq:tail-integral-full}
\end{align}
where $i$ and $j$ run over the atom types, $g_{ij}(r)$ is the corresponding pair
correlation function, $\rho_i$ and $\rho_j$ are the number densities of each
type, and $S(r)$ is the switching function that takes the
potential to zero.  This function must be continuous and take values $S(r_{in})
= 1$ and $S(r_{out}) = 0$; its derivative must also be continuous and take
values $S'(r_{in}) = S'(r_{out}) = 0$.

If we assume the pair correlation function $g_{ij}(r) \approx 1$ beyond $r =
r_\text{in}$ (which is usually a good approximation for liquids at relatively
large distances), the improper integral in the second term of
Equation~\eqref{eq:tail-integral-full} can be evaluated analytically for simple (e.g.
inverse-power) forms of the pair potential $\phi_{ij}(r)$.  Using a sixth-power
dispersion form $\phi_{ij}(r) = -C^6_{ij} r^{-6}$, the improper integral becomes
$\int_{r_\text{out}}^\infty 24\pi C^6_{ij} r^{-4} \dee r = 8\pi r_\text{out}^{-3}$
and we have:
\begin{equation}
    p_\text{exact} - p_\text{cut} \approx p_\text{corr} = \frac{1}{6} \sum_{i=1}^{n_\text{typ}} \rho_i
                                         \sum_{j=1}^{n_\text{typ}} \rho_j C^6_{ij}
                    \left[ \int_{r_\text{in}}^{r_\text{out}} - r \frac{\dee}{\dee r}
                        \left(\frac{1 - S(r)}{r^6}\right) 4\pi r^2 \dee r
                            + 8\pi r_\text{out}^{-3} \right] .
\label{eq:tail-integral-r6}
\end{equation}

Applying integration by parts to the remaining integral gives

\begin{equation}
    p_\text{corr} = \frac{1}{6} \sum_{i=1}^{n_\text{typ}} \rho_i
                                         \sum_{j=1}^{n_\text{typ}} \rho_j C^6_{ij}
                    \left[ -4\pi r_\text{out}^{-3} + \int_{r_\text{in}}^{r_\text{out}}
                        \left(\frac{1 - S(r)}{r^6}\right)  12 \pi r^2 \dee r
                            + 8\pi r_\text{out}^{-3} \right]
\label{eq:tail-integral-iparts}
\end{equation}

and simplifying and rearranging leaves us with

\begin{align}
    p_\text{corr} &= \frac{1}{6} \sum_{i=1}^{n_\text{typ}} \rho_i
                                         \sum_{j=1}^{n_\text{typ}} \rho_j C^6_{ij}
                    \left[ 4\pi r_\text{in}^{-3} - \int_{r_\text{in}}^{r_\text{out}}
                        12\pi r^{-4} S(r) \dee r \right]\nonumber\\
        &= \frac{2\pi}{3}  \sum_{i=1}^{n_\text{typ}} \rho_i
            \sum_{j=1}^{n_\text{typ}} \rho_j C^6_{ij} \left((1 - \lambda)r_\text{in}^{-3} + \lambda r_\text{out}^{-3}\right)
\label{eq:tail-integral-simplified}
\end{align}
with
\begin{equation*}
    \lambda = \frac{3\int_{r_\text{in}}^{r_\text{out}} r^{-4} S(r) \dee r}
        {r_\text{in}^{-3} - r_\text{out}^{-3}}
\end{equation*}

This form isolates the problematic integral $\int_{r_\text{in}}^{r_\text{out}}
r^{-4} S(r) \dee r$ which, depending on the form of the switching function
$S(r)$, may be complicated or impossible to do analytically.  For practical
simulations, however, $\lambda$ can simply be precomputed using any suitable
numerical method for a given value of $r_\text{in}$, $r_\text{out}$, and $S(r)$;
this value can be used throughout the simulation.

We can also see that Equation~\eqref{eq:tail-integral-simplified} takes the form
of a linear interpolation between the tail correction with a cutoff at
$r_\text{in}$ and the correction with a cutoff of $r_\text{out}$; if we choose
an $S(r)$ whose values are bounded between 0 and 1 then $\lambda$ will likewise
be bounded between 0 and 1.

On closer inspection, however, we can see that the interpolation endpoints are
\emph{not} the values the tail correction would take with a sharp cutoff:
If we start with Equation~\eqref{eq:tail-integral-r6} and take $S(r) = 1$ for $r <
r_\text{out}$, we get instead:

\begin{equation}
    p_\text{exact} - p_\text{cut,sharp} \approx \frac{4\pi}{3} \sum_{i=1}^{n_\text{typ}} \rho_i
                                         \sum_{j=1}^{n_\text{typ}} \rho_j C^6_{ij} r_\text{out}^{-3},
\label{eq:tail-integral-sharp}
\end{equation}

\begin{figure}
    \includegraphics[scale=0.8]{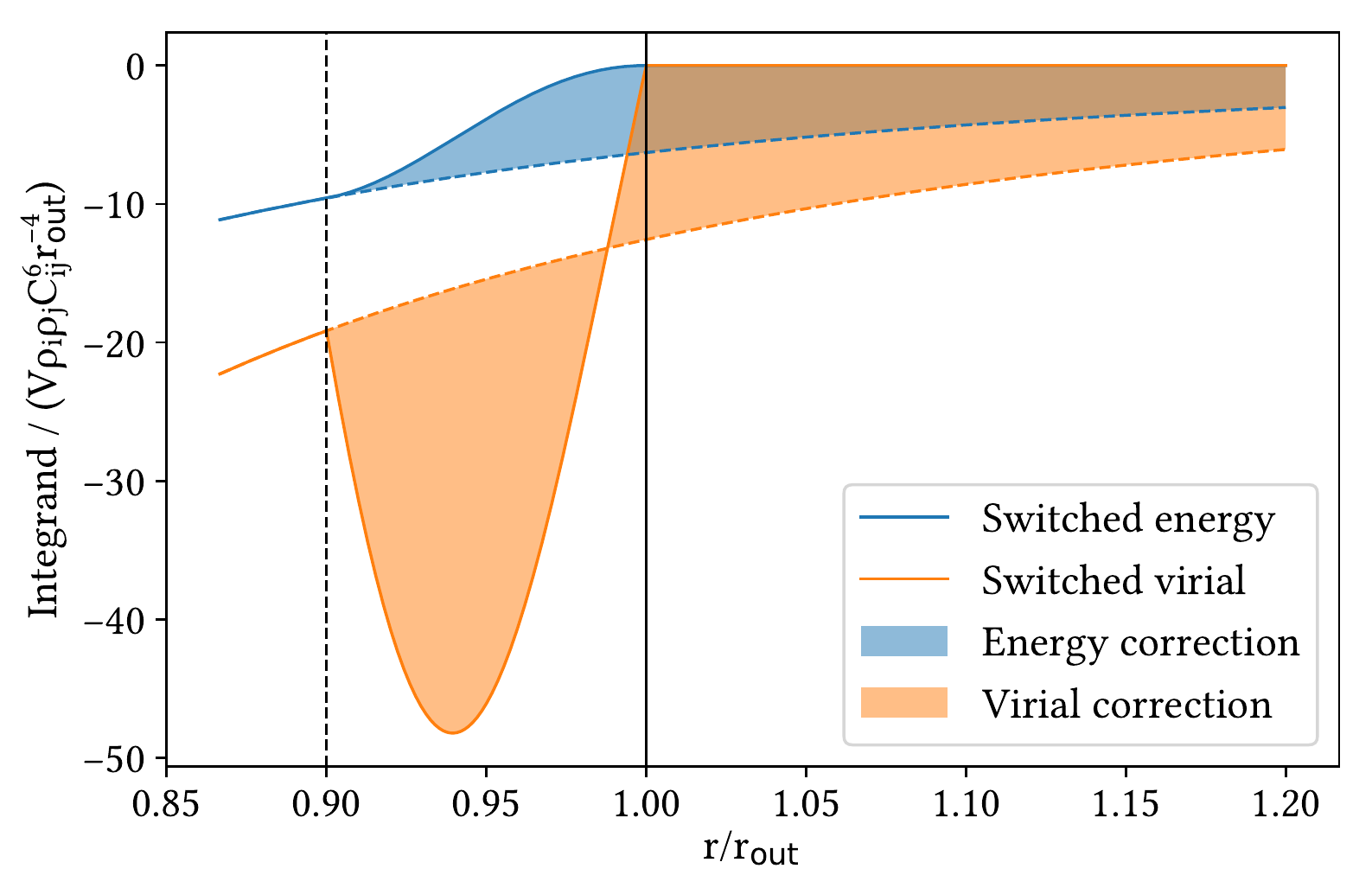}
    \caption{Illustration of the integrals for the energy and virial tail
        corrections for a potential of the form $\phi_\text{cut}(r_{ij}) =
        C^6_{ij}r_{ij}^{-6} S(r)$.  The dashed line corresponds to the potential
        with no cutoff ($f_E(r) = -\frac{1}{2} r^{-6} \cdot 4\pi r^2$ for the energy and $f_W(r) =
        \frac{1}{6} r \frac{\dee}{\dee r}r^{-6} \cdot 4\pi r^2$ for the virial), while the solid line
        corresponds to the potential multiplied by the switching function $S(r)$.
        The shaded areas between the dashed and solid lines depict the
        integrals of Equations~\eqref{eq:tail-integral-r6} (times $-V$)
        and~\eqref{eq:tail-integral-energy}; the area below the dotted line
        is negative.  The energy and virial integrals (areas) are equal.
    }
    \label{fig:tail-integrands}
\end{figure}

which is the correction Tildesley and Allen~\cite{Allen+1989} give for the
sixth-power part of an \lj\ potential, and \emph{twice} the value we would get from
Equation~\eqref{eq:tail-integral-simplified} by letting $\lambda = 1$.  This
discrepancy is due to the extra $-4\pi r_\text{out}^{-3}$ term that emerged from
the integration by parts in Equation~\eqref{eq:tail-integral-iparts}, and it can
be physically interpreted as follows: With a sharp cutoff, an atom feels no
force as it crosses the cutoff; the force just changes discontinuously from
$-\phi'(r\rightarrow r_\text{out}^-)$ to zero.  With a smooth cutoff, however, the switching
function provides an extra gentle inward force as the atom exits the transition
region.  The extra virial due to this force provides an \emph{effective} tail
correction to the system's overall pressure, albeit only about half of the
difference of the pressure with the sharp cutoff to the pressure of the ideal
system with an infinite cutoff; Figure~\ref{fig:tail-integrands} provides an
illustration of this idea.

Tail corrections may also be computed, using the same method as above, for the
energy.  Although they do not affect the simulation dynamics in any way, they
may be used for accurate bookkeeping and later analysis.  The expression
is~\cite{Allen+1989,Sun1998}
\begin{align}
    E_\text{exact} - E_\text{cut} &= \frac{V}{2} \sum_{i=1}^{n_\text{typ}} \rho_i
                                \sum_{j=1}^{n_\text{typ}} \rho_j
                    \int_{r_\text{in}}^\infty 
        (\phi_{ij}(r) - \phi_{ij,\text{cut}}(r)) 4 \pi r^2 g_{ij}(r) \dee r \nonumber\\
        &\approx \frac{V}{2} \sum_{i=1}^{n_\text{typ}} \rho_i
                                \sum_{j=1}^{n_\text{typ}} \rho_j C^6_{ij}
                    \left[\int_{r_\text{in}}^{r_\text{out}} -(1 - S(r)) 4 \pi r^{-4}  \dee r
                    + \int_{r_\text{out}}^\infty -4 \pi r^{-4} \dee r \right] \nonumber\\
        &= \frac{V}{2} \sum_{i=1}^{n_\text{typ}} \rho_i
                                \sum_{j=1}^{n_\text{typ}} \rho_j C^6_{ij}
                    \left[-\frac{4\pi}{3} r_\text{in}^{-3} +
                    \int_{r_\text{in}}^{r_\text{out}} 4\pi r^{-4}S(r) \dee r \right] \nonumber\\
        &= -\frac{2\pi V}{3} \sum_{i=1}^{n_\text{typ}} \rho_i
                                \sum_{j=1}^{n_\text{typ}} \rho_j C^6_{ij}
                    \left((1 - \lambda) r_\text{in}^{-3} + \lambda r_\text{out}^{-3}\right)
        \label{eq:tail-integral-energy}
\end{align}
with $\lambda$ defined as before.  This is -- coincidentally, with the $r^{-6}$
potential -- identical to the virial correction, i.e. the pressure correction
in Equation~\eqref{eq:tail-integral-simplified} multiplied by $-V$.

\bibliographystyle{aipnum4-1}
\bibliography{refs}